\documentclass[a4paper,UKenglish,cleveref, autoref, thm-restate]{lipics-v2021}
\usepackage{todonotes}
\usepackage{cite}
\usepackage{amsmath,amssymb,amsfonts}
\usepackage{rotating}
\usepackage{algorithmic}
\usepackage{graphicx}
\usepackage{threeparttable}
\usepackage{textcomp}
\usepackage{xcolor}

\newcommand{\review}[1]{\textcolor{black}{#1}}

\usepackage{url}

\def\BibTeX{{\rm B\kern-.05em{\sc i\kern-.025em b}\kern-.08em
    T\kern-.1667em\lower.7ex\hbox{E}\kern-.125emX}}

\usepackage{booktabs, multirow}
\usepackage{pifont}
\usepackage{colortbl} 

\definecolor{verylightgray}{rgb}{0.93, 0.93, 0.93}

\usepackage[skins]{tcolorbox}

\usepackage{tcolorbox}
\tcbuselibrary{skins}
\newenvironment{summarybox}
{\begin{tcolorbox}
[enhanced,arc=0mm,colback=gray!10,frame hidden,overlay unbroken={%
    \draw[thick,black] (interior.north west)--(interior.south west);
},left=2pt,right=0pt,top=0pt,bottom=0pt,before={\vspace{3pt}\noindent},after={\vspace{0pt}}]}
{\end{tcolorbox}}
\newenvironment{summary}
{\vspace{5pt}\noindent\begin{summarybox}}
{\end{summarybox}\vspace{-5pt}}



\category{Technical Track Paper}

\nolinenumbers 

\supplement{}

\supplementdetails[subcategory={Replication Package and Appendices}, cite={anonymous_replication_2026}, swhid={}]{Software}{https://doi.org/10.5281/zenodo.21971227}

\title{Green AI: Cost of LLM-Based Code Completion}

\author{Negar Alizadeh}{Utrecht University, Utrecht, The Netherlands}{n.s.alizadeh@uu.nl}{https://orcid.org/0009-0005-8155-165X}{}
\author{Nishant Saurabh}{Utrecht University, Utrecht, The Netherlands}{n.saurabh@uu.nl}{https://orcid.org/0000-0002-1926-4693}{}
\author{Fernando Castor}{University of Twente, Enschede, The Netherlands}{f.castor@utwente.nl}{https://orcid.org/0000-0002-6389-3630}{}

\authorrunning{N. Alizadeh, N. Saurabh and F.Castor}

\Copyright{Negar Alizadeh, Nishant Saurabh and Fernando Castor} 

\EventEditors{Robert Feldt, Maria Paasivaara, Daniel Mendez, Stefan Wagner, and Marvin Mu\~{n}oz Bar\'{o}n}
\EventNoEds{5}
\EventLongTitle{20th International Symposium on Empirical Software Engineering and Measurement (ESEM 2026)}
\EventShortTitle{ESEM 2026}
\EventAcronym{ESEM}
\EventYear{2026}
\EventDate{October 8--9, 2026}
\EventLocation{Munich, Germany}
\EventLogo{}
\SeriesVolume{394}
\ArticleNo{42}

\begin{document}

\maketitle

\begin{abstract}
\textbf{Background:} Code completion is one of the most widely used applications of large language models (LLMs) in software development. In addition to proprietary coding assistants, powerful open-weight LLMs are increasingly adopted for locally deployed code completion systems, partly motivated by privacy concerns. Despite advances in LLM accuracy, the energy cost of inference in code completion tasks remains underexplored, particularly under large-context workloads and across programming languages. 
\textbf{Aims:} This study investigates the trade-off between accuracy and energy consumption in LLM-based code completion and analyzes how workload characteristics, context size, and model scale influence inference energy usage.
\textbf{Method:} We evaluate \review{25} open-weight LLMs on two complementary code completion workloads: repository-level next-line completion (left-to-right prediction) with varying context sizes on the RepoBench dataset, and fill-in-the-middle (FIM) code completion across Python, Java, and Rust on the McEval dataset. We further analyze the influence of input tokens, output tokens, model size, and their interactions on energy consumption using correlation analysis and cluster-robust linear regression models.
\textbf{Results}: Our findings show that the dominant drivers of energy consumption depend strongly on the structure of the completion task. In RepoBench, energy consumption is primarily influenced by input context size and its interaction with model scale, whereas in McEval, output generation and its interaction with active parameter count become the dominant factors. We further observe that output generation is substantially more energy-intensive per token than prompt processing. Across both benchmarks, smaller and heavily quantized models frequently achieve Pareto-optimal trade-offs, often providing accuracy comparable to larger FP16 models while consuming substantially less energy.
\textbf{Conclusions}: The energy behavior of LLM inference in software engineering tasks depends strongly on workload structure, context length, and model scale. Our results suggest that increasing model size or context length does not necessarily lead to proportionally better completion quality, while quantization can substantially improve energy efficiency with limited accuracy degradation. These findings contribute toward more energy-aware deployment strategies for sustainable AI-assisted software development.

\end{abstract}
\section{Introduction}
\label{sec:intro}
Since the emergence of large language models (LLMs) and their rapid advancement in software development tasks, these models have become increasingly popular among software companies and individual programmers, \review{with 84\% of developers using or planning to use AI tools~\cite{stack2025survey}.}
Furthermore, numerous IDEs now incorporate LLM-based code completion tools (e.g., GitHub Copilot, JetBrains AI Assistant, Gemini Code Assist, Claude Code, and Tabnine) and many AI-powered coding environments such as Cursor and Devin are emerging.


LLMs are known to be energy-intensive. Prior work has examined the energy consumption of language models~\cite{luccioni2022estimating,Patternson:2022:CFM} with earlier work focusing mostly on the training phase~\cite{luccioni2022estimating,luccioni2024power} and more recent papers also investigating the costs of inference~\cite{zschache2025comparing, adamska2025green,fernandez-etal-2025-energy, poddar-etal-2025-towards}. Training is orders of magnitude more expensive, but inference grows with scale and can represent the majority of the energy cost of LLMs for long-term energy use. \review{Prior measurements report that approximately three-fifths of ML energy use is attributed to inference and two-fifths to training~\cite{Patternson:2022:CFM}.} Despite this, most previous work has focused on natural language tasks, with few studies investigating the energy efficiency of LLMs in software development. Existing studies of inference energy indicate that energy usage varies across tasks~\cite{alizadeh2025language,luccioni2024power}. This highlights the importance of evaluating LLM performance and energy efficiency not only for natural language tasks but also for other domains such as programming. Even within coding tasks, performance and energy consumption may vary across programming languages due to differences in syntax complexity, ecosystem maturity, and training data availability. For example, Python's syntax is less verbose than Java's, with the former being dynamically typed and the latter statically typed. Furthermore, Rust has a unique type system among mainstream programming languages and comparatively less training data. 

This study investigates the energy use of LLMs in code completion tasks. 
Previous studies on LLM inference energy focused primarily on output-intensive tasks, such as code generation and bug fixing~\cite{alizadeh2025language}, and input tokens are often considered relatively inexpensive in comparison~\cite{fernandez2025energy,solovyeva2026towards,pizzini2026sweetspot}. However, in long-context software engineering workloads such as repository-level code completion, processing large input contexts can become a major energy bottleneck. In scenarios where input contexts consist of thousands of tokens while the output contains only a few lines, the relative contribution of input processing becomes significantly larger. This study aims to quantify the relative impact of input and output tokens across different code completion workloads.
We focus our investigation on locally-deployable, open weight models. The ability to run these models locally enables us to obtain precise energy measurement while controlling for interference from external factors. The architecture of smaller models often mirrors that of larger ones and results are to some extent transferrable.

Our study examines two real-world code completion scenarios: (i) \textbf{Repository-level next-line completion}, where cross-file context is also included in the prompt; and (ii) \textbf{Infill code completion}, where the model fills in missing code using both preceding and succeeding context. 
We further investigate LLM performance and energy consumption across multiple programming languages, namely Python, Java, and Rust, with the goal of uncovering deeper insights into the energy–performance dynamics of LLMs in realistic coding tasks.
We address the following research questions, aiming to contribute to more energy-aware development and deployment strategies for AI-assisted programming tools:
\vspace{0.2cm}
\begin{description}
\item[RQ1] What is the trade-off between energy consumption and accuracy in code completion?
\begin{description}
\item[RQ1.1] How does the trade-off differ across programming languages?
\item[RQ1.2] How does this trade-off vary across different context sizes?
\end{description}
\item[RQ2] What is the relative influence of input and output token counts on total energy consumption, and how do these drivers interact with model scale?
\end{description}

\section{Research Method}
\label{sec:method}
\begin{figure}[tb]
\centerline{
\includegraphics[width=0.9\linewidth]
{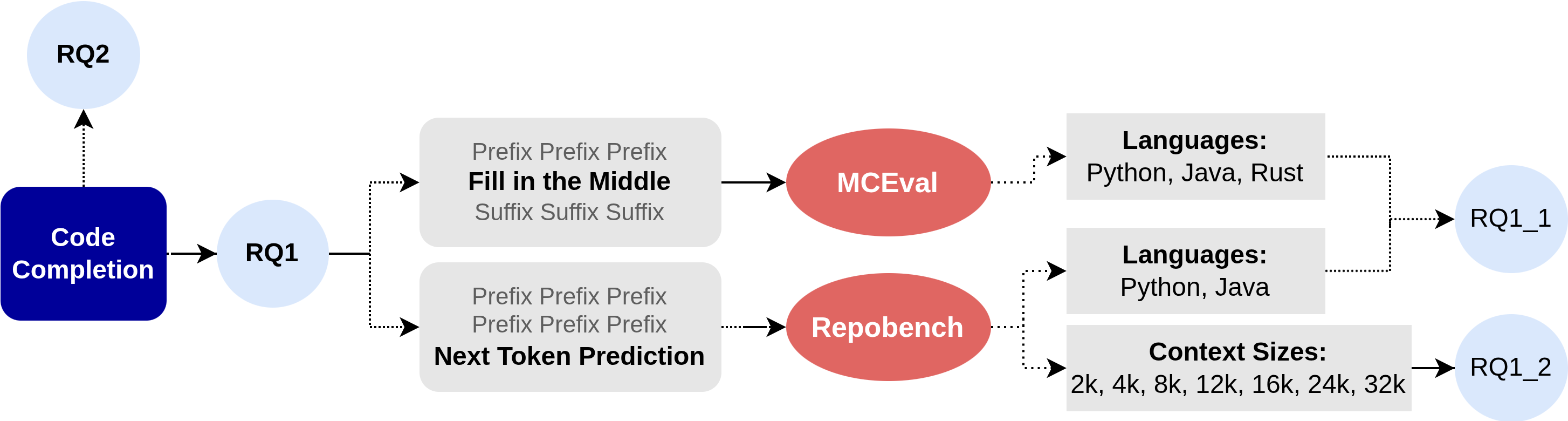}}
\caption{Schematic representation of the study.}
\label{fig1}
\end{figure}

\textbf{Figure~\ref{fig1}} presents a schematic outline of our study along two directions, corresponding to the two primary categories of code completion~\cite{gong2024eval}: Left-to-Right (L2R) prediction (aka. next token prediction) and Fill-in-the-Middle (FIM) prediction (aka. infilling).
\\
Left-to-Right (L2R) completion refers to the traditional autoregressive setting (one token at a time) in which the model continues the code solely based on the preceding context. This paradigm is commonly used in repository-level completion tasks, where the model may receive large amounts of contextual information from multiple files, functions, and project-wide dependencies. As a result, L2R completion often involves long-context workloads that require processing thousands of input tokens before generating the prediction.
In contrast, Fill-in-the-Middle (FIM) completion predicts missing code between a given prefix and suffix. This setting better reflects interactive editing scenarios in modern IDEs, where developers modify or insert code within existing files rather than only appending new content at the end.
Our experimental setup is specifically designed to address each category using the appropriate methodology, as described in the remainder of this section.

\subsection{Execution Environment}

\textbf{To run LLMs locally}, an LLM inference engine is required. Several open-source LLM inference engines have recently emerged to enable local deployment and inference of open-weight LLMs, \review{including PrivateGPT (57.4K GitHub stars), GPT4All (77.4K), llamafile (25.5K), vLLM (87.6K), Ollama (177K) at the time of this study. We selected Ollama (v0.13.0) due to its popularity, as reflected by its GitHub star metric.} It is particularly useful for its support of multiple quantization levels, enabling efficient execution on commodity hardware. Furthermore, its seamless model-switching capability and extensive API features make it well-suited for exploration and reproducible experiments. Additionally, its strong adoption among developers, as reported in the 2025 Stack Overflow Developer Survey\cite{stack2025survey}, further reinforces the external validity of our tool selection.

To monitor GPU utilization and \textbf{to measure power consumption}, as suggested in previous studies\cite{alizadeh2024green, alizadeh2025language, argerich2024measuring, castor2024estimating}, we use the Python interface (nvidia-ml-py v13.590.48) for the NVIDIA Management Library (NVML) \cite{nvidia_nvml}. For every running model, power usage is sampled every 100ms, and total energy consumption is estimated as the arithmetic mean of the recorded power samples multiplied by the elapsed execution time.
While higher sampling frequencies yield a more detailed power profile, we adopt a 10Hz sampling rate, \review{consistent with prior work~\cite{alizadeh2025language}}, to avoid the runtime overhead associated with more frequent measurements.
The reported power values represent net consumption, computed by subtracting the idle power of the GPU from the measured power. To measure GPU idle power, we executed the nvidia-smi\cite{nvidia_smi} command for 340 seconds, collecting 3400 samples at 100 ms intervals. 3400 collected observations provide a 98\% confidence level that the true power consumption lies within ±2\% of the observed values. The exact command we have used is provided in the replication package~\cite{anonymous_replication_2026}. The mean measured idle power of the GPU was 46.82 W. 

All experiments were conducted on a cluster equipped with an NVIDIA A100 80GB PCIe GPU~\cite{nvidia_a100_product_brief_2022} configured in ``High Performance'' mode, two AMD 7313 CPUs running at 3\,GHz with 1\,TB of system memory and 32\,MB cache, and AlmaLinux 8.10 (64-bit) operating system configured with the ``Performance'' CPU governor. Each experiment was repeated three times to mitigate transient fluctuations, and the reported values represent the average across the three runs. No other jobs were allowed to run on the GPU during the experiments.

\subsection{Dataset}\label{sec:dataset}
\textbf{Repobench}~\cite{liu2023repobench} is designed for repository-level auto code completion. We used the code completion subset, which focuses on next-line prediction using three distinct settings: Cross-File-First simulates predicting code that hasn't yet appeared in the current file. This is the hardest task that requires LLMs to know about the external files. In Cross-File-Random cases, a random and non-first occurrence of a cross-file line is masked, so the code has already been partially introduced in the file, allowing the LLMs to leverage prior in-file context. Finally, the In-File split contains cases within a single file, where an in-file line unrelated to cross-file modules is masked.
Compared to other repository-level benchmarks for code completion~\cite{zhang2023repocoder, microsoft_lcc_python, li2024evocodebench, le2025impacts, cheng2024fullstack, ding2023crosscodeeval}, Repobench supports a wide range of context sizes, varying from 2000 to 128000 input tokens. This makes it particularly suitable for our study, as it allows us to better investigate the effect of input length on energy consumption. Considering the limitations imposed by the context length of the LLMs under examination, we were only able to carry out experiments on the data with a maximum of 32000 as their context limit.
Then, for each context level, 100 samples were extracted for each split (300), resulting in 2100 code completion problems.
\\
\textbf{McEval}~\cite{chai2024mceval} is a multilingual code evaluation benchmark, covering 40 languages.
Its code completion task can be categorized into single-line completion, multi-line completion, and span completion. In single-line completion, only one line of code is masked, whereas in the other two settings multiple lines are masked. In multi-line completion, the masked lines are dispersed throughout the code, while in span completion they appear consecutively as a single block. To evaluate LLM performance for FIM purposes across different programming languages, we selected Python because it is the most widely used language for machine learning development, Java as a representative of popular compiled languages, and Rust as a less common but still mainstream language, reflecting scenarios where training data is relatively limited. The selected McEval subsets contain 340 Python, 355 Java, and 343 Rust completion tasks. While other languages such as C\# or C++, could also be included, we deliberately limited our selection to these three for practical reasons. This choice can be acknowledged as a potential threat to validity.

\subsection{Model Selection}

To ensure diversity in both model scale and architecture, we collected models ranging from approximately 1B up to 20B parameters\review{, focusing on model sizes practical for local deployment and inference.} The list of models under evaluation covers multiple parameter ranges, including models below 4B parameters, several models in the 6B--8B range, and larger models from 13B to 20B parameters.
To construct this set, we selected model families based on their availability in GGUF format and their popularity on the Ollama platform, as indicated by pull counts. Two models (Granite 3.3 8B and Granite 3.3 2B) were not available on Ollama in full-precision and Q8 formats; therefore, we sourced their GGUF files directly from the official IBM Granite repository on Hugging Face~\cite{ibm_granite_2024}.
\\
We also included newer models with efficiency-driven approaches, as they are relevant to the energy-efficiency goals of this study. In particular, we included both dense and Mixture-of-Experts (MoE) architectures. Unlike dense models, where all parameters participate during inference, MoE models activate only a subset of their parameters for each token, potentially reducing inference cost while maintaining strong performance~\cite{Rajbhandari2022DeepSpeedMoEAM, grootendorst2024visual}. Therefore, throughout this study, we report the number of active parameters for MoE models rather than the total parameter count.
To investigate the trade-off between numerical precision, accuracy, and energy consumption, we evaluate models under multiple quantization levels, including FP16, Q8, and Q4, \review{representing 16-bit floating-point, 8-bit, and 4-bit weight precision, respectively}. \textbf{Quantization} reduces the numerical precision used to represent model weights, lowering memory usage and computational overhead during inference~\cite{jin2024comprehensive, gong2024survey, del2025aggregating}.
\\
In line with our focus on code completion, we only include general-purpose and coding-specific models. For experiments using the RepoBench dataset, base versions of the models are required, whereas instruction-tuned variants are needed for the McEval dataset.
\\
Base LLMs are pretrained models optimized for next-token prediction, whereas instruction-tuned are additionally trained to follow natural language instructions.
Following the original benchmark papers, RepoBench~\cite{liu2023repobench} requires base models because the benchmark is designed to evaluate the completion of the next-line code at the repository-level in an autoregressive setting, where the model must continue the code directly from the provided context without relying on the following behavior of the instruction. In contrast, McEval~\cite{chai2024mceval} follows a prompt-based fill-in-the-middle setting, making instruction-tuned variants more appropriate for the evaluation scenario.
Moreover, in the RepoBench setting, context length is particularly important because models with larger context windows can better capture long-range dependencies.
We also considered the reputation and maturity of model developers, as models from established AI organizations tend to be better maintained and widely adopted.
\review{A complete overview of the evaluated models is provided in the online appendix~\cite{anonymous_replication_2026}.}

\subsection{Performance metrics} \label{performance metrics}
There are numerous metrics to evaluate the correctness of the code generated by LLMs. This section focuses on the key metrics frequently used to assess LLM performance. Given that this research cannot explore all possible metrics, it focuses on the metrics most frequently used to determine the performance of LLMs used for code completion.
\review{A recent systematic literature review of LLM-based code completion~\cite{husein_large_2025} identified Top-$k$ Accuracy, Levenshtein Edit Similarity (ES), Exact Match (EM), Mean Reciprocal Rank (MRR), and BLEU as the most commonly used evaluation metrics.}
For the RepoBench dataset, to ensure comparability and reproducibility, we use the same evaluation metrics as defined in the original RepoBench paper~\cite{liu2023repobench}: Normalized Edit Similarity and Exact Match.
Although these metrics may appear overly strict, since functionally correct code can differ in naming conventions, they are appropriate in a repository-level setting. In such scenarios, the model is provided with extensive contextual information, including cross-file references and project-wide naming patterns. Therefore, we expect the LLM to generate a piece of code that is not only functionally correct but also adheres to the intended naming conventions for variables, functions, and classes. Moreover, even when the generated code is not fully correct, high textual similarity can still be practically useful in software development, as developers may adapt and refine the generated code instead of writing it entirely from scratch.

For the generated code in the McEval dataset, we utilized the Docker container prepared by the dataset authors to compute Pass@1 accuracy based on the provided test cases.
\review{This metric corresponds to the Plausible@1 criterion recommended in recent LLM evaluation guidelines~\cite{guideline2026}.}
Since McEval consists of relatively small and self-contained examples, it is feasible to execute the generated code and verify correctness through automated testing. All necessary environments are preinstalled in the Docker image, and the correctness is assessed by executing the generated code against the provided tests~\cite{chai2024mceval}.
This approach is better suited for this use case, where no cross-file repository context is involved, ensuring that the evaluation reflects functional correctness rather than syntactic similarity.

In contrast, applying execution-based evaluation to RepoBench would require resolving repository dependencies and building complete software systems, which would substantially limit the scale and reproducibility of this study. Therefore, for RepoBench, we rely on similarity-based metrics instead. This approach is better suited for the repository-level setting while still allowing us to evaluate how well the generated code aligns with the expected implementation and project-wide conventions.

\subsection{Experimental Procedure}
To ensure consistency, we used identical hyperparameters for all models within each dataset. The temperature and top-p values were set to 0.2 and 0.95, respectively, as prior studies~\cite{chen2021evaluating, liu2023repobench} suggest these settings yield better performance in code generation tasks.
\\
For \textbf{RepoBench}, the maximum number of output tokens was set to 128, following the configuration of the original study. Only the context size was varied to accommodate different input lengths. To construct the prompts, we used the \textit{construct\_prompt} function provided with the dataset.
For \textbf{McEval}, we first measured the length of the longest canonical solution for each language (Python: 1227 tokens, Java: 1274 tokens, Rust: 1017 tokens). We then set the maximum number of output tokens to 1300 to ensure coverage of the longest solutions across all three languages. The context size remained at its default value of 2000 tokens. As for the prompt, we used the \textit{instruction} column provided in the dataset.

\section{Results}
\label{sec:res}
In this section, we present our findings following the order of the research questions. 
\subsection{RQ1-What is the trade-off between energy consumption and accuracy in code completion?}
To address this RQ, we first collected the energy consumption of all LLMs while executing the complete workload. For McEval, we computed the mean Pass@1 score and total energy consumption across the complete workload of each language subset. Since both the size and the content of the workloads differ across subsets, we report the energy consumed per generated token (J/token) to enable a fair comparison between models.
The same consideration applies to RepoBench, where the Python and Java subsets also contain different problems. Similar to McEval, we report aggregated energy consumption values together with the mean Edit Similarity scores. The results are presented in Tables~\ref{tab:mc_res} and~\ref{tab:repo_res}. Due to page limitations, only the FP16 results are included in the main paper, while the results for 8-bit and 4-bit quantizations are provided in the online appendix~\cite{anonymous_replication_2026}.
\review{To analyze the trade-off between accuracy and energy consumption, we employ the Pareto frontier, a multi-objective optimization method that identifies nondominated configurations across competing objectives.}

\begin{table*}[!htp]
\centering
\caption{Results of inference for instruction-tuned models in Python, Java and Rust on McEval. Total energy (E) is in Wh, elapsed time (T) is in seconds, J/t indicates the energy per generated token and P@1 indicates pass@1. Note that the workload across languages is not the same.}
\label{tab:mc_res}
\scriptsize
\setlength{\tabcolsep}{3pt}
\renewcommand{\arraystretch}{0.95}

\resizebox{\textwidth}{!}{%
\begin{tabular}{lcccccccccccc}
\toprule
\multirow{2}{*}{Model} &
\multicolumn{4}{c}{Python} &
\multicolumn{4}{c}{Java} &
\multicolumn{4}{c}{Rust} \\
\cmidrule(lr){2-5}
\cmidrule(lr){6-9}
\cmidrule(lr){10-13}
& T & E & J/t & P@1 & T & E & J/t & P@1 & T & E & J/t & P@1 \\
\midrule
\cellcolor[HTML]{E8E8E8}codegemma:7b & \cellcolor[HTML]{E8E8E8}997.86 & \cellcolor[HTML]{E8E8E8}66.32 & \cellcolor[HTML]{E8E8E8}3.51 & \cellcolor[HTML]{E8E8E8}68.18 & \cellcolor[HTML]{E8E8E8}898.07 & \cellcolor[HTML]{E8E8E8}59.42 & \cellcolor[HTML]{E8E8E8}3.53 & \cellcolor[HTML]{E8E8E8}66.76 & \cellcolor[HTML]{E8E8E8}898.13 & \cellcolor[HTML]{E8E8E8}59.41 & \cellcolor[HTML]{E8E8E8}3.53 & \cellcolor[HTML]{E8E8E8}63.26 \\
codellama:13b & 1288.67 & 88.94 & 4.98 & 56.97 & 1293.52 & 89.56 & 4.98 & 54.65 & 1357.33 & 94.07 & 4.98 & 53.35 \\
\cellcolor[HTML]{E8E8E8}codellama:7b & \cellcolor[HTML]{E8E8E8}839.38 & \cellcolor[HTML]{E8E8E8}57.79 & \cellcolor[HTML]{E8E8E8}2.75 & \cellcolor[HTML]{E8E8E8}50.30 & \cellcolor[HTML]{E8E8E8}997.59 & \cellcolor[HTML]{E8E8E8}68.85 & \cellcolor[HTML]{E8E8E8}2.75 & \cellcolor[HTML]{E8E8E8}55.49 & \cellcolor[HTML]{E8E8E8}934.90 & \cellcolor[HTML]{E8E8E8}64.58 & \cellcolor[HTML]{E8E8E8}2.74 & \cellcolor[HTML]{E8E8E8}49.56 \\
deepseek-coder-v2:16b & 582.85 & 30.98 & 1.58 & 73.64 & 603.88 & 32.54 & 1.58 & \textbf{89.30} & 596.67 & 31.48 & 1.58 & \textbf{78.13} \\
\cellcolor[HTML]{E8E8E8}deepseek-coder:1.3b & \cellcolor[HTML]{E8E8E8}847.85 & \cellcolor[HTML]{E8E8E8}48.53 & \cellcolor[HTML]{E8E8E8}0.74 & \cellcolor[HTML]{E8E8E8}5.76 & \cellcolor[HTML]{E8E8E8}805.12 & \cellcolor[HTML]{E8E8E8}46.16 & \cellcolor[HTML]{E8E8E8}0.73 & \cellcolor[HTML]{E8E8E8}7.61 & \cellcolor[HTML]{E8E8E8}883.53 & \cellcolor[HTML]{E8E8E8}50.69 & \cellcolor[HTML]{E8E8E8}0.73 & \cellcolor[HTML]{E8E8E8}3.79 \\
deepseek-coder:6.7b & 1264.01 & 87.28 & 2.77 & 66.06 & 1394.03 & 96.45 & 2.75 & 64.79 & 1254.37 & 86.60 & 2.75 & 55.98 \\
\cellcolor[HTML]{E8E8E8}deepseek-llm:7b & \cellcolor[HTML]{E8E8E8}773.20 & \cellcolor[HTML]{E8E8E8}51.26 & \cellcolor[HTML]{E8E8E8}2.76 & \cellcolor[HTML]{E8E8E8}35.15 & \cellcolor[HTML]{E8E8E8}831.19 & \cellcolor[HTML]{E8E8E8}55.43 & \cellcolor[HTML]{E8E8E8}2.76 & \cellcolor[HTML]{E8E8E8}33.24 & \cellcolor[HTML]{E8E8E8}851.87 & \cellcolor[HTML]{E8E8E8}56.89 & \cellcolor[HTML]{E8E8E8}2.75 & \cellcolor[HTML]{E8E8E8}19.53 \\
gemma3n:e4b-it & 993.42 & 38.42 & 2.53 & 58.18 & 1052.29 & 41.54 & 2.49 & 25.07 & 1030.64 & 41.94 & 2.49 & 41.98 \\
\cellcolor[HTML]{E8E8E8}gemma:2b & \cellcolor[HTML]{E8E8E8}398.66 & \cellcolor[HTML]{E8E8E8}24.54 & \cellcolor[HTML]{E8E8E8}1.25 & \cellcolor[HTML]{E8E8E8}25.15 & \cellcolor[HTML]{E8E8E8}383.24 & \cellcolor[HTML]{E8E8E8}23.45 & \cellcolor[HTML]{E8E8E8}1.25 & \cellcolor[HTML]{E8E8E8}25.92 & \cellcolor[HTML]{E8E8E8}346.91 & \cellcolor[HTML]{E8E8E8}20.97 & \cellcolor[HTML]{E8E8E8}1.25 & \cellcolor[HTML]{E8E8E8}20.12 \\
gemma:7b & 1850.42 & 125.69 & 3.55 & 43.94 & 1718.55 & 116.72 & 3.51 & 47.04 & 1744.44 & 118.70 & 3.48 & 34.98 \\
\cellcolor[HTML]{E8E8E8}granite-code:20b & \cellcolor[HTML]{E8E8E8}2416.59 & \cellcolor[HTML]{E8E8E8}166.91 & \cellcolor[HTML]{E8E8E8}7.75 & \cellcolor[HTML]{E8E8E8}64.24 & \cellcolor[HTML]{E8E8E8}2006.78 & \cellcolor[HTML]{E8E8E8}138.60 & \cellcolor[HTML]{E8E8E8}7.63 & \cellcolor[HTML]{E8E8E8}65.92 & \cellcolor[HTML]{E8E8E8}2003.26 & \cellcolor[HTML]{E8E8E8}138.45 & \cellcolor[HTML]{E8E8E8}7.63 & \cellcolor[HTML]{E8E8E8}63.26 \\
granite-code:3b & 605.66 & 36.32 & 1.88 & 51.82 & 455.01 & 27.07 & 1.87 & 36.34 & 479.93 & 28.42 & 1.88 & 40.23 \\
\cellcolor[HTML]{E8E8E8}granite-code:8b & \cellcolor[HTML]{E8E8E8}940.17 & \cellcolor[HTML]{E8E8E8}64.20 & \cellcolor[HTML]{E8E8E8}3.53 & \cellcolor[HTML]{E8E8E8}65.15 & \cellcolor[HTML]{E8E8E8}813.09 & \cellcolor[HTML]{E8E8E8}55.48 & \cellcolor[HTML]{E8E8E8}3.53 & \cellcolor[HTML]{E8E8E8}69.58 & \cellcolor[HTML]{E8E8E8}800.20 & \cellcolor[HTML]{E8E8E8}54.54 & \cellcolor[HTML]{E8E8E8}3.53 & \cellcolor[HTML]{E8E8E8}67.64 \\
granite3-moe:3b & 663.20 & 25.38 & 0.92 & 39.39 & 446.74 & 16.95 & 0.93 & 30.70 & 644.38 & 24.49 & 0.92 & 33.53 \\
\cellcolor[HTML]{E8E8E8}granite3.3:2b & \cellcolor[HTML]{E8E8E8}454.22 & \cellcolor[HTML]{E8E8E8}27.30 & \cellcolor[HTML]{E8E8E8}1.36 & \cellcolor[HTML]{E8E8E8}47.58 & \cellcolor[HTML]{E8E8E8}469.16 & \cellcolor[HTML]{E8E8E8}27.99 & \cellcolor[HTML]{E8E8E8}1.37 & \cellcolor[HTML]{E8E8E8}39.72 & \cellcolor[HTML]{E8E8E8}516.69 & \cellcolor[HTML]{E8E8E8}30.90 & \cellcolor[HTML]{E8E8E8}1.36 & \cellcolor[HTML]{E8E8E8}35.86 \\
granite3.3:8b & 707.74 & 47.96 & 3.38 & 40.61 & 820.77 & 55.99 & 3.39 & 58.03 & 829.16 & 56.49 & 3.39 & 47.52 \\
\cellcolor[HTML]{E8E8E8}llama3:8b & \cellcolor[HTML]{E8E8E8}612.80 & \cellcolor[HTML]{E8E8E8}38.92 & \cellcolor[HTML]{E8E8E8}3.12 & \cellcolor[HTML]{E8E8E8}50.30 & \cellcolor[HTML]{E8E8E8}697.16 & \cellcolor[HTML]{E8E8E8}44.67 & \cellcolor[HTML]{E8E8E8}3.12 & \cellcolor[HTML]{E8E8E8}58.03 & \cellcolor[HTML]{E8E8E8}706.18 & \cellcolor[HTML]{E8E8E8}45.47 & \cellcolor[HTML]{E8E8E8}3.12 & \cellcolor[HTML]{E8E8E8}56.27 \\
mistral:7b & 985.96 & 68.04 & 2.85 & 44.54 & 1070.87 & 73.95 & 2.84 & 38.59 & 972.74 & 67.09 & 2.85 & 37.61 \\
\cellcolor[HTML]{E8E8E8}phi3:14b & \cellcolor[HTML]{E8E8E8}1347.70 & \cellcolor[HTML]{E8E8E8}93.26 & \cellcolor[HTML]{E8E8E8}5.13 & \cellcolor[HTML]{E8E8E8}54.85 & \cellcolor[HTML]{E8E8E8}1375.09 & \cellcolor[HTML]{E8E8E8}95.39 & \cellcolor[HTML]{E8E8E8}5.13 & \cellcolor[HTML]{E8E8E8}65.92 & \cellcolor[HTML]{E8E8E8}1323.67 & \cellcolor[HTML]{E8E8E8}91.80 & \cellcolor[HTML]{E8E8E8}5.13 & \cellcolor[HTML]{E8E8E8}50.73 \\
phi3:3.8b & 694.29 & 47.46 & 1.83 & 34.85 & 857.20 & 58.86 & 1.82 & 30.70 & 844.41 & 57.92 & 1.83 & 18.37 \\
\cellcolor[HTML]{E8E8E8}qwen2.5-coder:1.5b & \cellcolor[HTML]{E8E8E8}373.94 & \cellcolor[HTML]{E8E8E8}16.64 & \cellcolor[HTML]{E8E8E8}1.04 & \cellcolor[HTML]{E8E8E8}56.97 & \cellcolor[HTML]{E8E8E8}345.69 & \cellcolor[HTML]{E8E8E8}15.47 & \cellcolor[HTML]{E8E8E8}1.04 & \cellcolor[HTML]{E8E8E8}65.07 & \cellcolor[HTML]{E8E8E8}352.92 & \cellcolor[HTML]{E8E8E8}15.40 & \cellcolor[HTML]{E8E8E8}1.04 & \cellcolor[HTML]{E8E8E8}54.81 \\
qwen2.5-coder:7b & 493.23 & 31.16 & 3.10 & \textbf{88.18} & 429.44 & 26.60 & 3.12 & 29.86 & 460.24 & 28.83 & 3.12 & 40.82 \\
\cellcolor[HTML]{E8E8E8}qwen3:4b & \cellcolor[HTML]{E8E8E8}520.68 & \cellcolor[HTML]{E8E8E8}31.44 & \cellcolor[HTML]{E8E8E8}2.17 & \cellcolor[HTML]{E8E8E8}76.06 & \cellcolor[HTML]{E8E8E8}465.15 & \cellcolor[HTML]{E8E8E8}28.00 & \cellcolor[HTML]{E8E8E8}2.17 & \cellcolor[HTML]{E8E8E8}68.17 & \cellcolor[HTML]{E8E8E8}531.88 & \cellcolor[HTML]{E8E8E8}32.49 & \cellcolor[HTML]{E8E8E8}2.16 & \cellcolor[HTML]{E8E8E8}70.26 \\
starcoder2:15b & 1503.64 & 103.40 & 6.02 & 76.36 & 1516.90 & 104.56 & 6.06 & 81.97 & 1571.24 & 108.33 & 6.10 & 69.68 \\
\bottomrule
\end{tabular}%
}
\end{table*}
\begin{figure}[!htb]
\centering

\begin{minipage}[b]{0.47\textwidth}
\includegraphics[width=\textwidth]{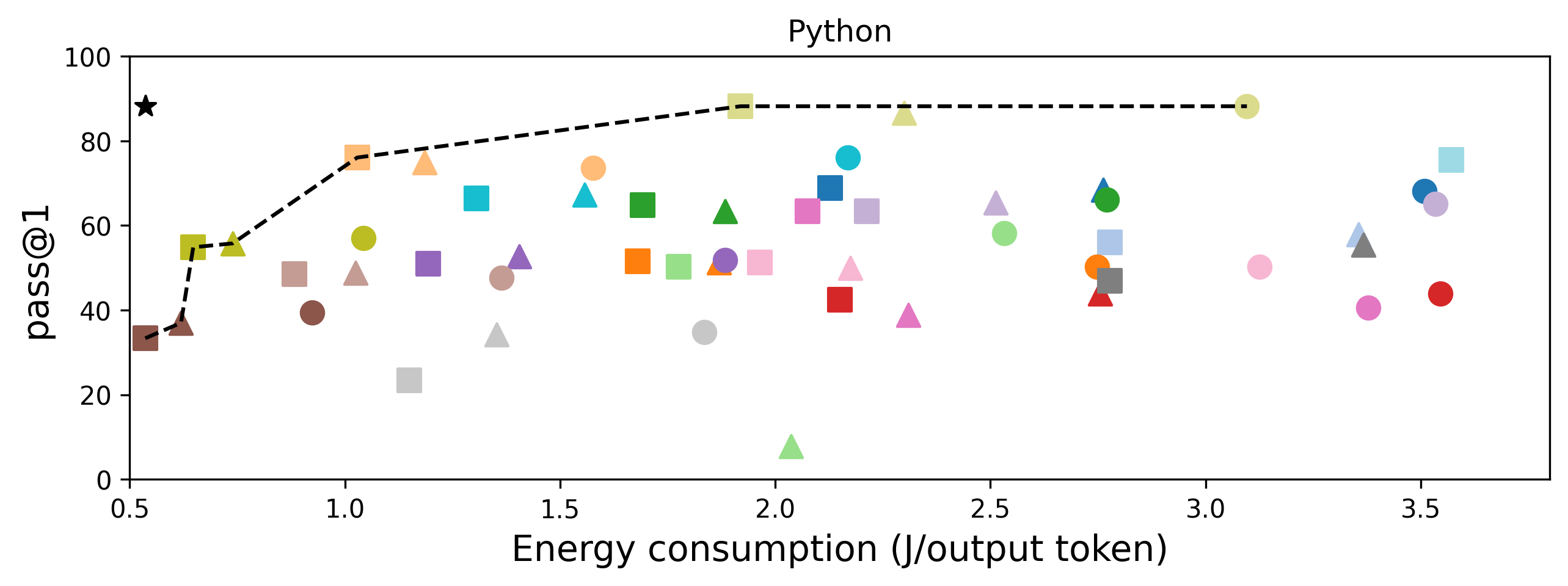}
\end{minipage}
\hspace{0.01\textwidth}
\begin{minipage}[b]{0.47\textwidth}
\includegraphics[width=\textwidth]{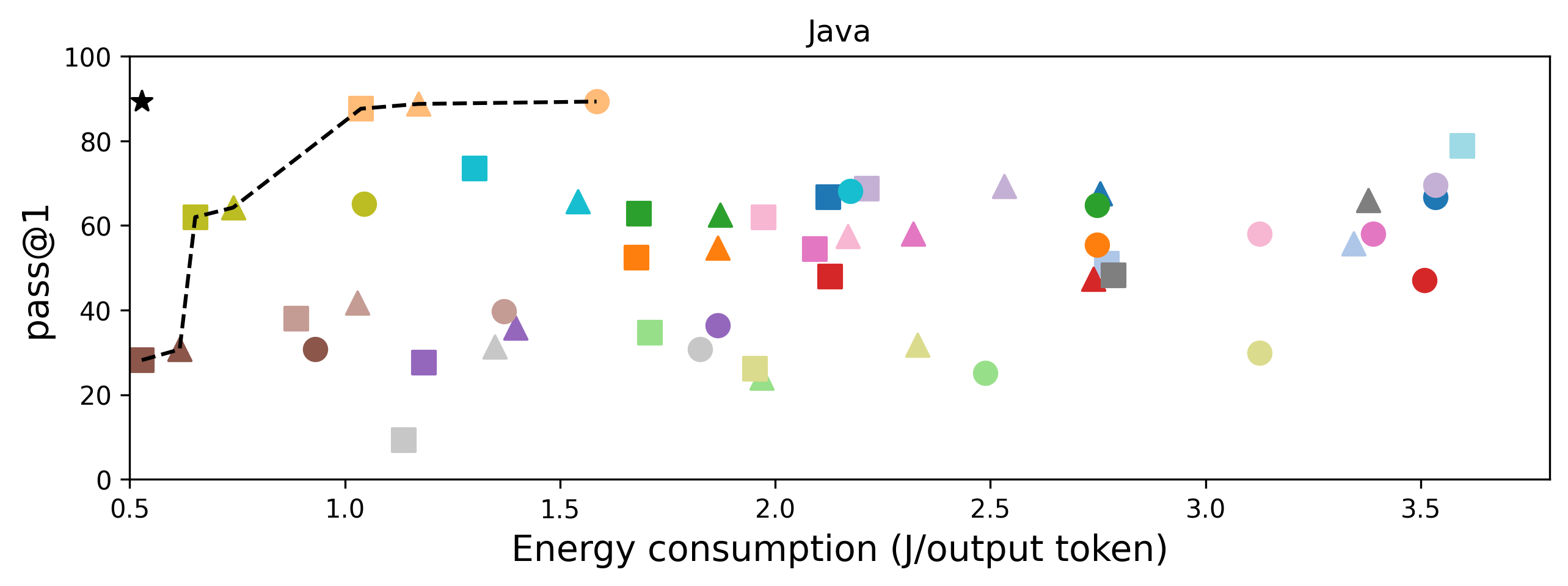}

\end{minipage}

\vspace{0.2cm}
\begin{minipage}[b]{0.47\textwidth}
\includegraphics[width=\textwidth]{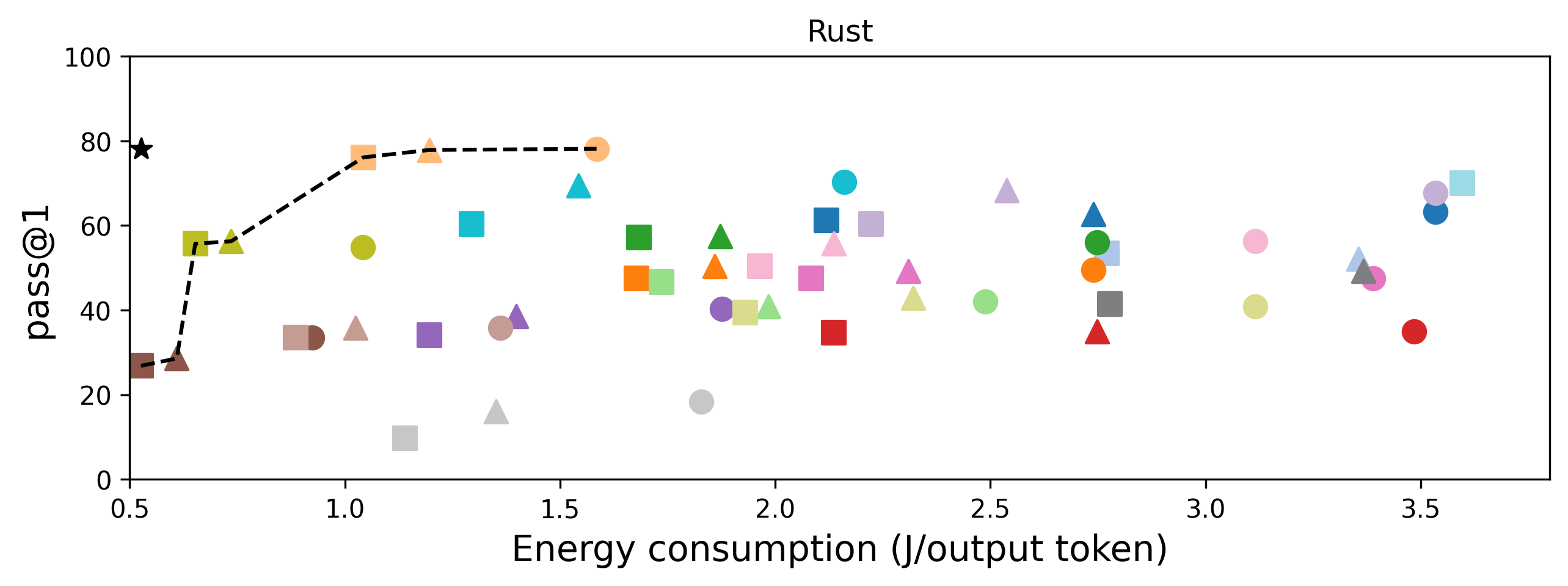}

\end{minipage}
\begin{minipage}[b]{0.47\textwidth}
\raisebox{0.5cm}{
\includegraphics[width=\textwidth]{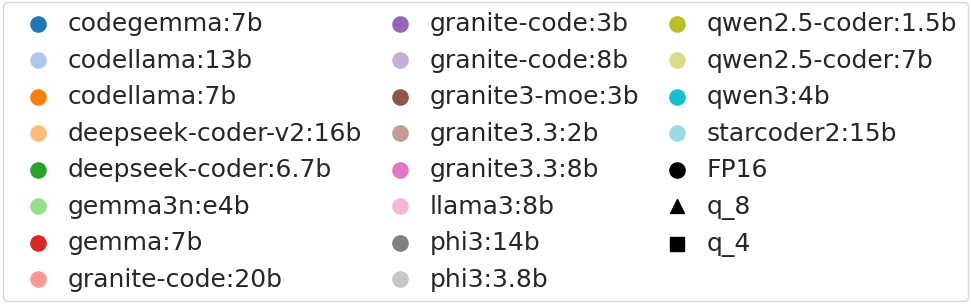}}
\end{minipage}
\caption{Relation between accuracy and energy consumption (Joule/token) for different languages in McEval. The black dashed line indicates the Pareto frontier and \textbf{\ding{72}} denotes the ideal point.}
\label{fig:mc_pareto}
\end{figure}
\begin{figure*}[!htb]
    \centering
    \begin{minipage}[b]{0.49\textwidth}
        \centering
        \includegraphics[width=\textwidth]{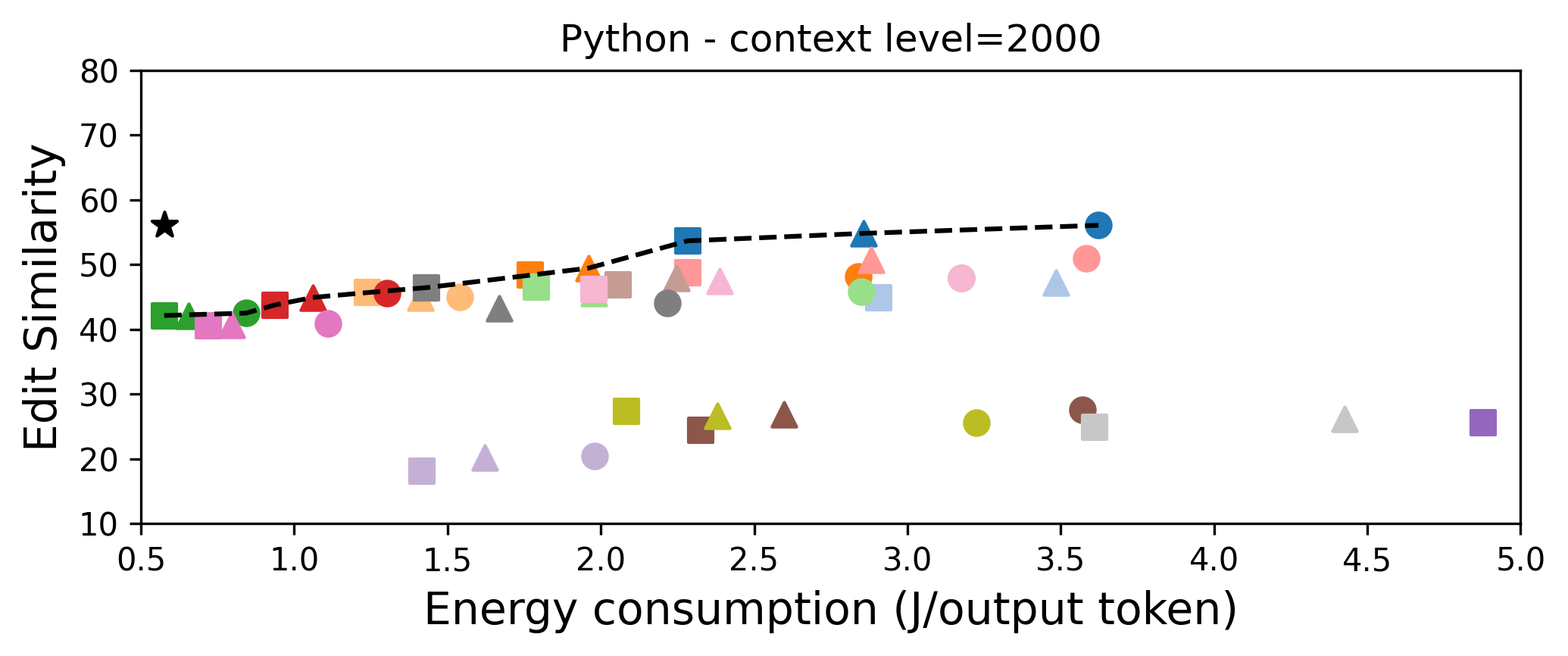}
    \end{minipage}
    \hfill
    \begin{minipage}[b]{0.49\textwidth}
        \centering
        \includegraphics[width=\textwidth]{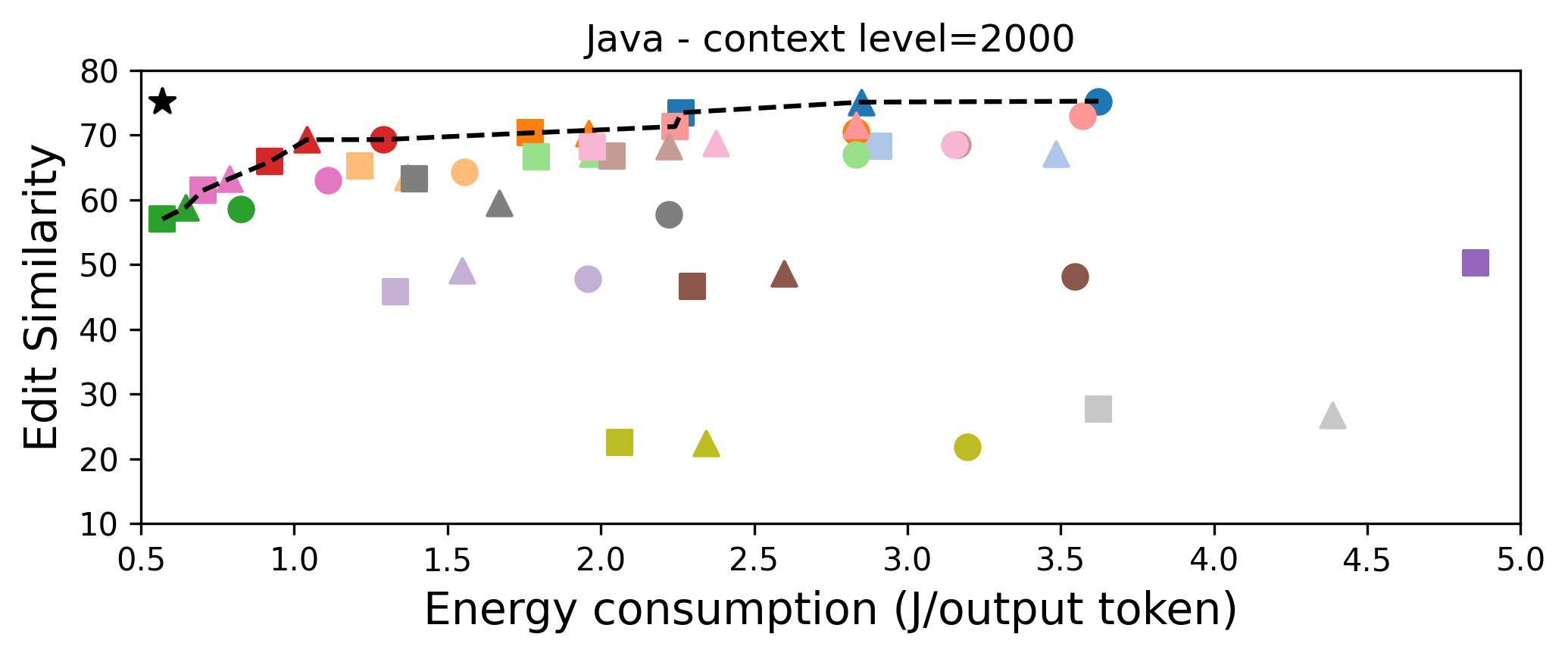}
    \end{minipage}
    
    \vspace{0.1cm}
    
    \begin{minipage}[b]{0.49\textwidth}
        \centering
        \includegraphics[width=\textwidth]{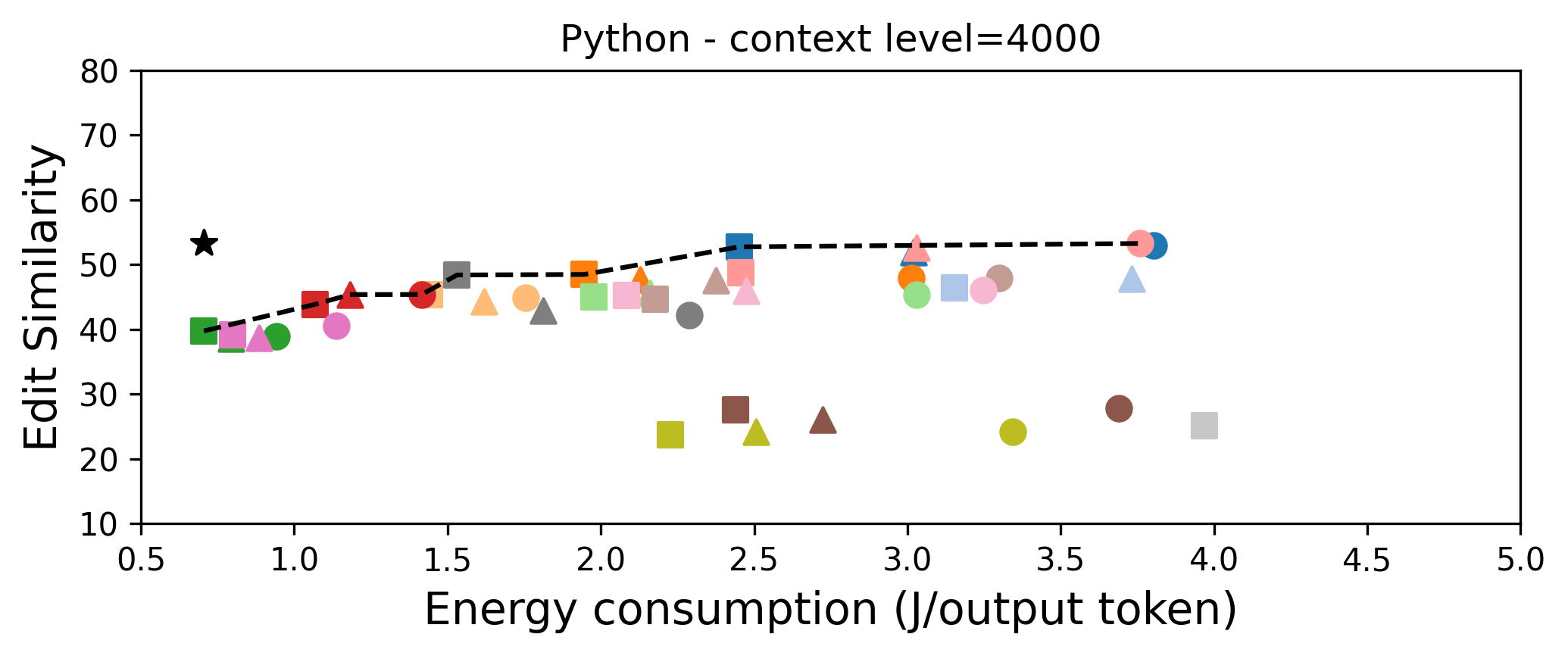}
    \end{minipage}
    \hfill
    \begin{minipage}[b]{0.49\textwidth}
        \centering
        \includegraphics[width=\textwidth]{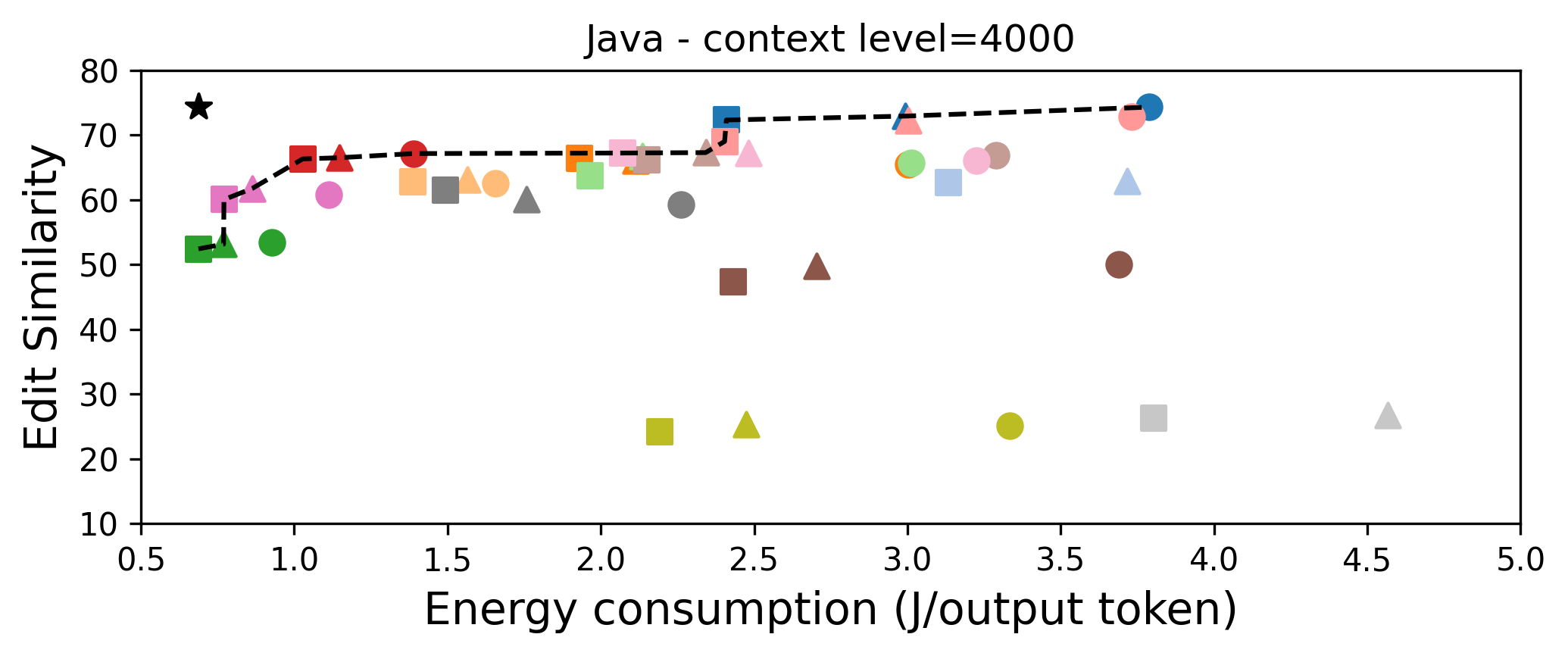}
    \end{minipage}
    \vspace{0.1cm}
    
    \begin{minipage}[b]{0.49\textwidth}
        \centering
        \includegraphics[width=\textwidth]{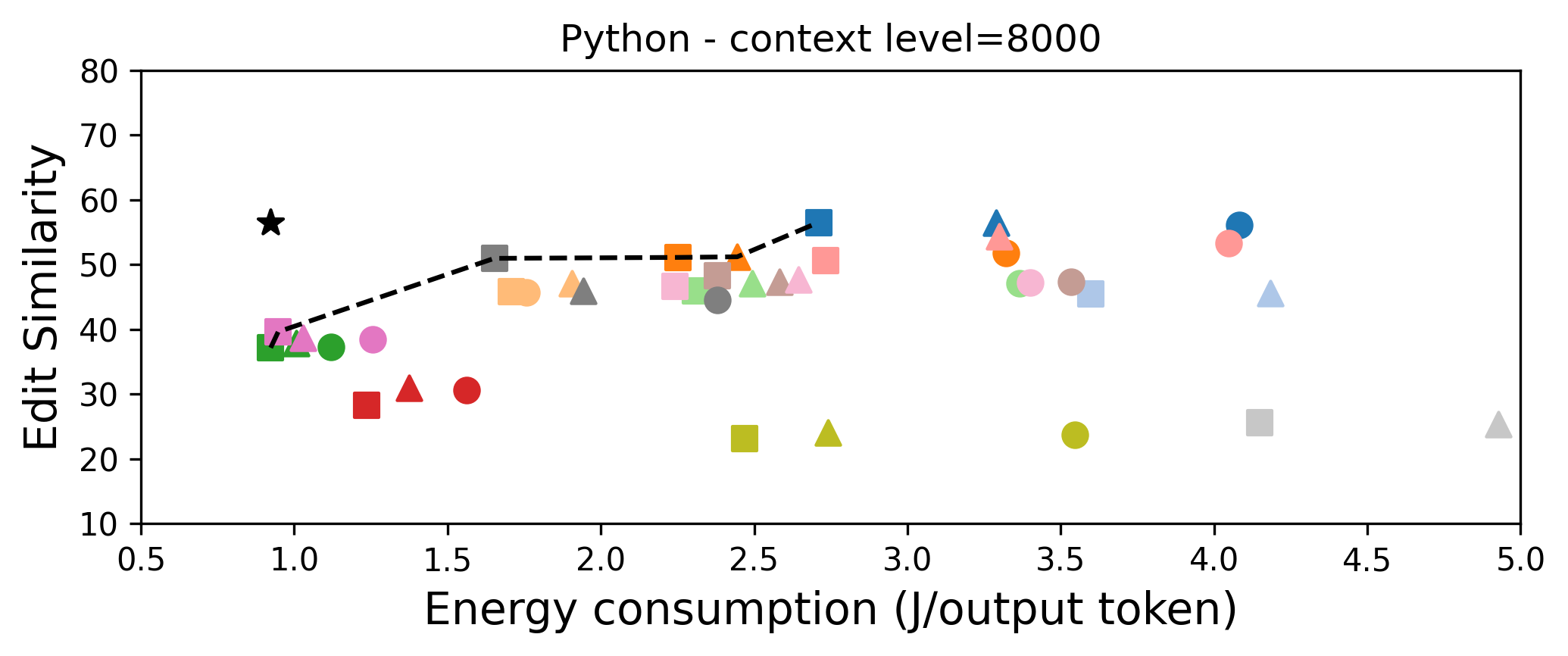}
    \end{minipage}
    \hfill
    \begin{minipage}[b]{0.49\textwidth}
        \centering
        \includegraphics[width=\textwidth]{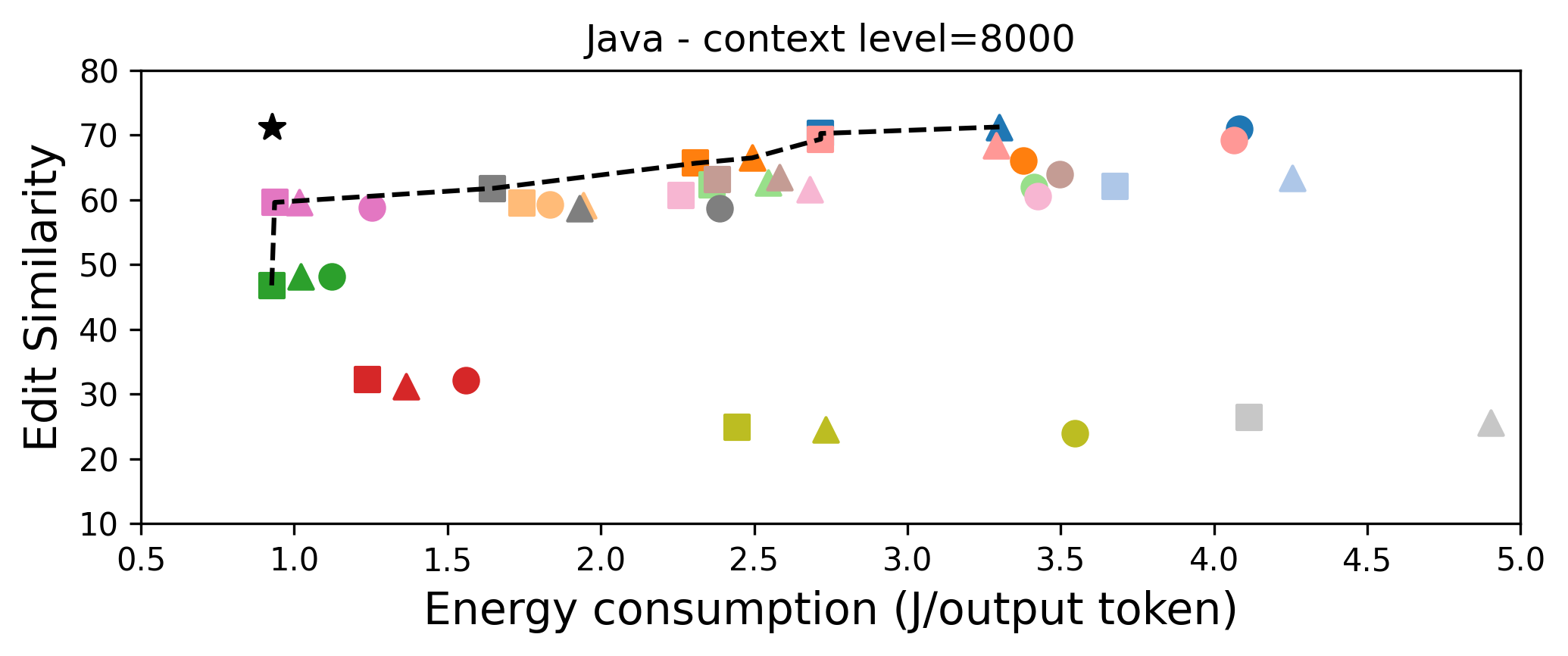}
    \end{minipage}
\vspace{0.1cm}
    
    \begin{minipage}[b]{0.49\textwidth}
        \centering
        \includegraphics[width=\textwidth]{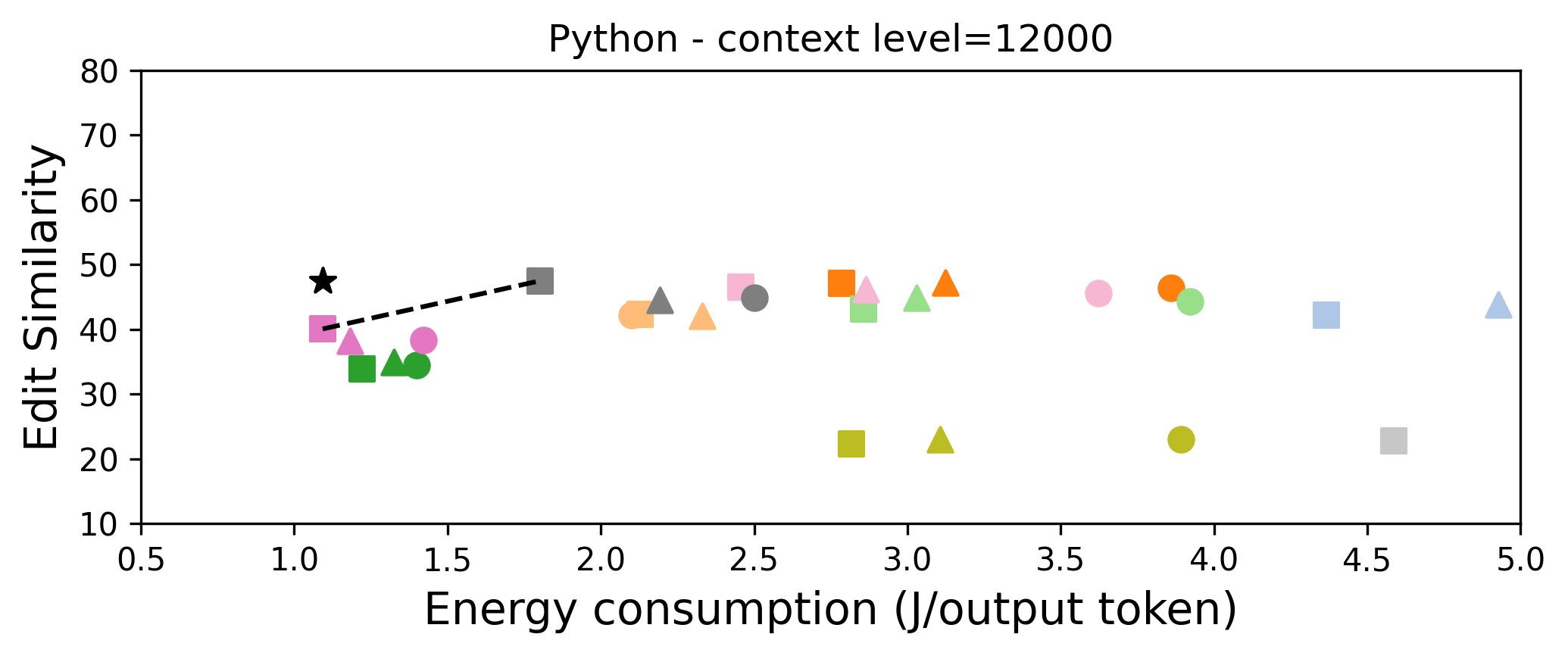}
    \end{minipage}
    \hfill
    \begin{minipage}[b]{0.49\textwidth}
        \centering
        \includegraphics[width=\textwidth]{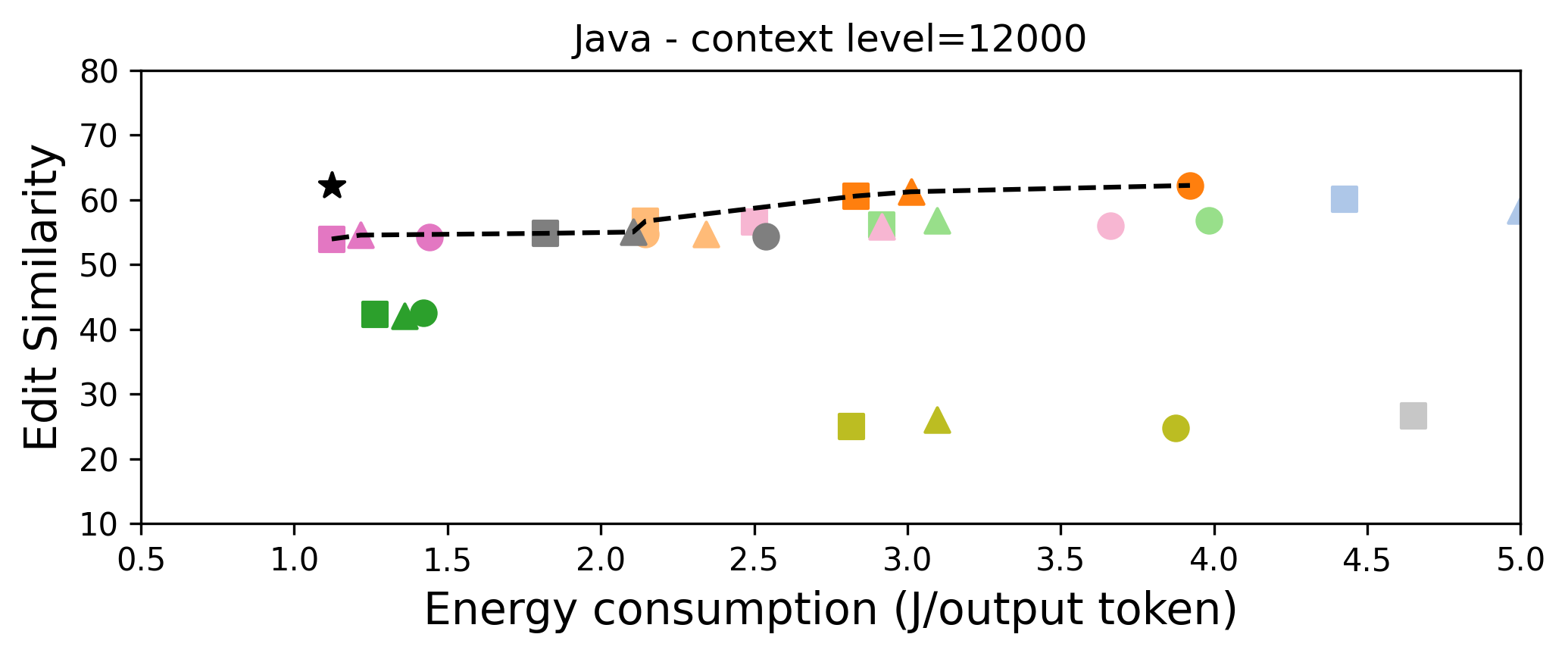}
    \end{minipage}
\vspace{0.1cm}
    
    \begin{minipage}[b]{0.49\textwidth}
        \centering
        \includegraphics[width=\textwidth]{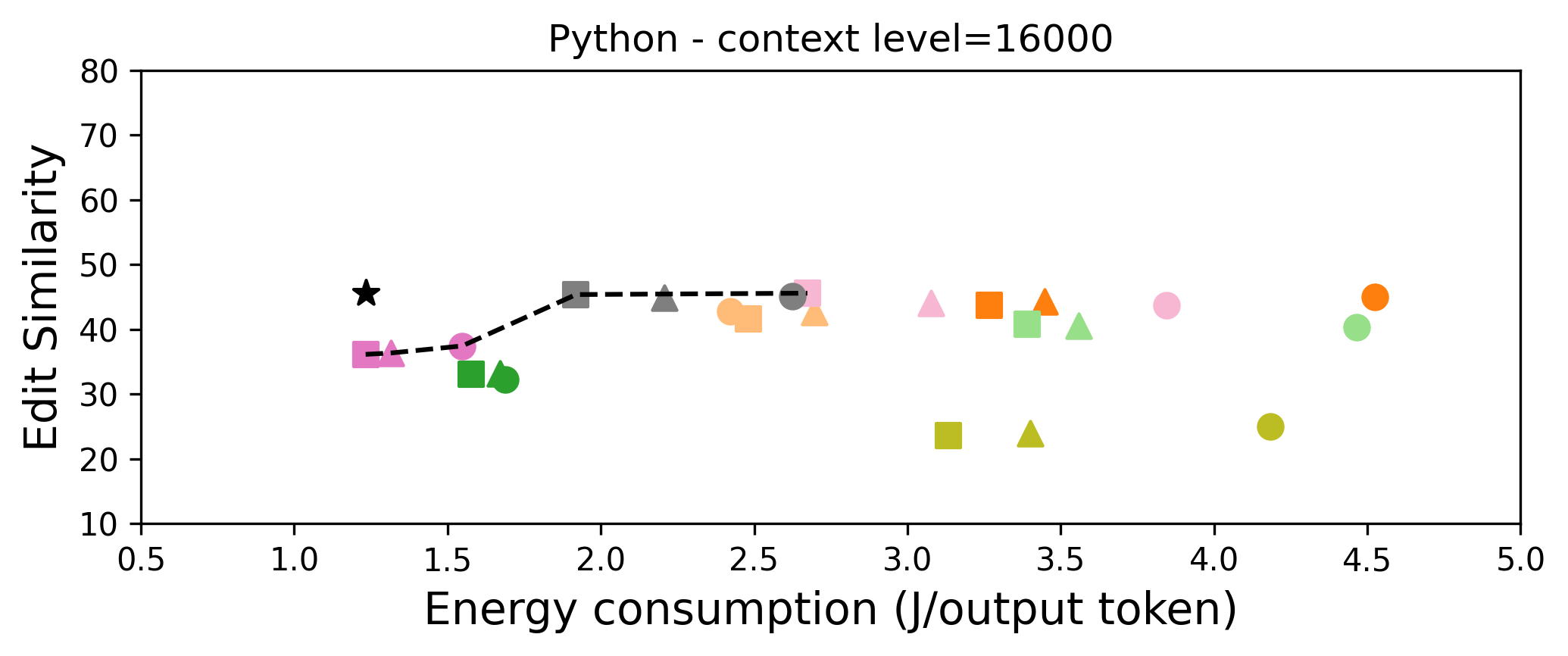}
    \end{minipage}
    \hfill
    \begin{minipage}[b]{0.49\textwidth}
        \centering
        \includegraphics[width=\textwidth]{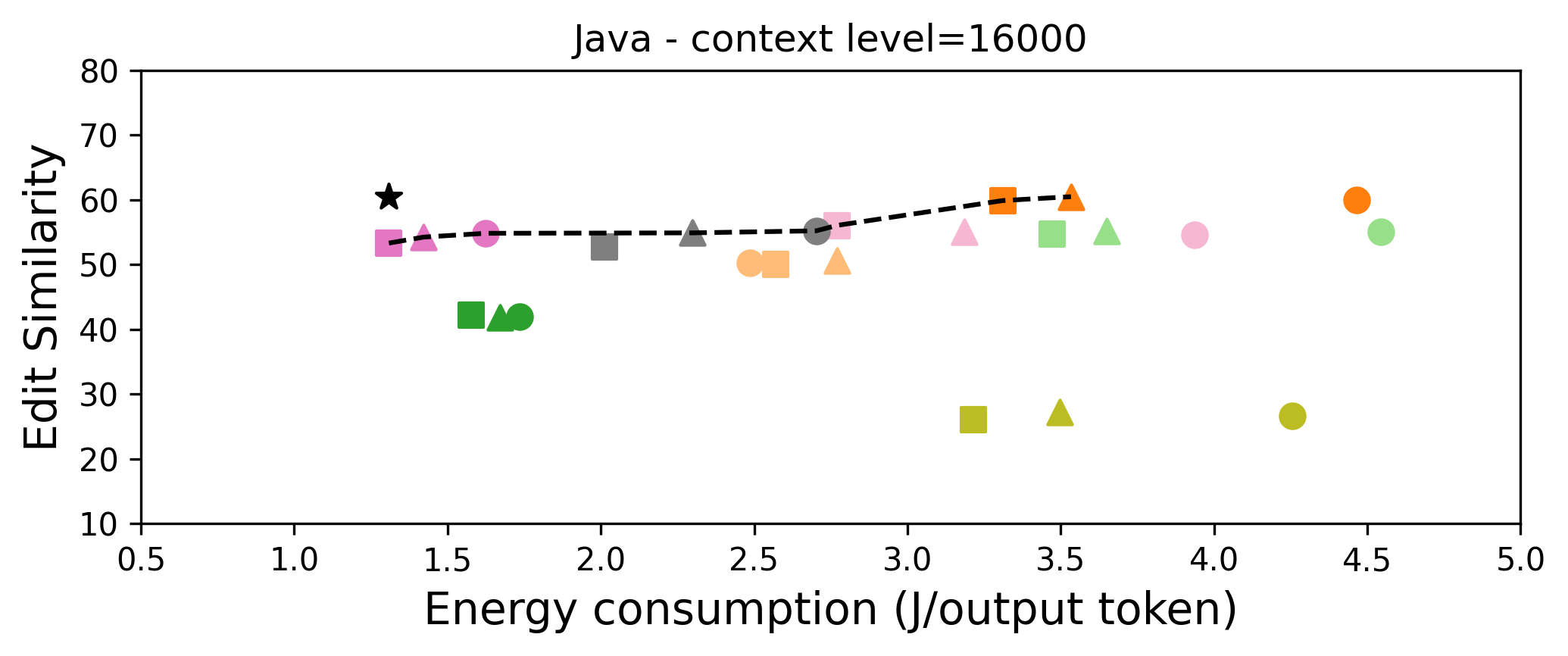}
    \end{minipage}

    \begin{minipage}[b]{1\textwidth}
        \centering
        \includegraphics[width=1\textwidth]{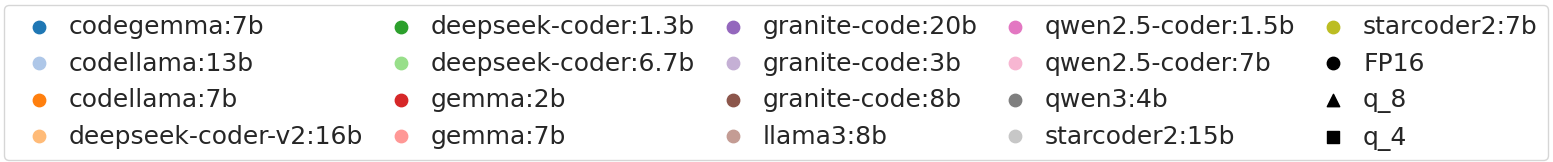}
    \end{minipage}
    
    \caption{Relation between accuracy and energy per generated token (Joule/token) for different context levels in Repobench dataset. The left column is the results for generating Python code, and the right column depicts the results for generating Java code. The black dashed line indicates the Pareto frontier. The \textbf{\ding{72}} denotes the ideal point. (Data points exceeding the x-axis limit are not displayed.)}
\label{fig:repo_pareto}
\end{figure*}
\begin{table*}[!htp]\centering
\begin{threeparttable}
\caption{Results of inference for base models in Python and Java on RepoBench. Total energy is reported in Wh, elapsed time in seconds, J/t indicates the energy per generated token and ES stands for Edit Similarity. The full list of models and Exact Match results is provided in the online appendix~\cite{anonymous_replication_2026}. Note that the workload across context sizes and languages is not the same.}\label{tab:repo_res}
\vspace{-3mm}
\scriptsize
\begin{tabular}{clcccccccc}
\toprule
\multirow{2}{*}{ctx} & \multirow{2}{*}{Model name} & \multicolumn{4}{c}{Python} & \multicolumn{4}{c}{Java} \\
\cmidrule(lr){3-6}
\cmidrule(lr){7-10}
& & Time & Energy & J/t & ES & Time & Energy & J/t & ES \\
\midrule
\multirow{17}{*}{2k} & \cellcolor[HTML]{E8E8E8}codegemma:7b & \cellcolor[HTML]{E8E8E8}624.62 & \cellcolor[HTML]{E8E8E8}40.16 & \cellcolor[HTML]{E8E8E8}3.62 & \cellcolor[HTML]{E8E8E8}\textbf{56.08} & \cellcolor[HTML]{E8E8E8}640.51 & \cellcolor[HTML]{E8E8E8}41.26 & \cellcolor[HTML]{E8E8E8}3.62 & \cellcolor[HTML]{E8E8E8}\textbf{75.25} \\
 & codellama:13b & 885.98 & 60.45 & 5.10 & 47.86 & 894.54 & 61.08 & 5.10 & 67.36 \\
 & \cellcolor[HTML]{E8E8E8}codellama:7b & \cellcolor[HTML]{E8E8E8}506.89 & \cellcolor[HTML]{E8E8E8}34.16 & \cellcolor[HTML]{E8E8E8}2.84 & \cellcolor[HTML]{E8E8E8}48.12 & \cellcolor[HTML]{E8E8E8}506.04 & \cellcolor[HTML]{E8E8E8}34.05 & \cellcolor[HTML]{E8E8E8}2.83 & \cellcolor[HTML]{E8E8E8}70.58 \\
 & deepseek-coder-v2:16b & 432.11 & 19.88 & 1.54 & 44.97 & 403.57 & 18.72 & 1.56 & 64.36 \\
 & \cellcolor[HTML]{E8E8E8}deepseek-coder:1.3b & \cellcolor[HTML]{E8E8E8}153.59 & \cellcolor[HTML]{E8E8E8}8.68 & \cellcolor[HTML]{E8E8E8}0.84 & \cellcolor[HTML]{E8E8E8}42.51 & \cellcolor[HTML]{E8E8E8}153.84 & \cellcolor[HTML]{E8E8E8}8.52 & \cellcolor[HTML]{E8E8E8}0.83 & \cellcolor[HTML]{E8E8E8}58.55 \\
 & deepseek-coder:6.7b & 450.25 & 29.96 & 2.85 & 45.87 & 434.74 & 28.66 & 2.83 & 67.06 \\
 & \cellcolor[HTML]{E8E8E8}gemma:2b & \cellcolor[HTML]{E8E8E8}256.03 & \cellcolor[HTML]{E8E8E8}14.67 & \cellcolor[HTML]{E8E8E8}1.30 & \cellcolor[HTML]{E8E8E8}45.55 & \cellcolor[HTML]{E8E8E8}254.84 & \cellcolor[HTML]{E8E8E8}14.65 & \cellcolor[HTML]{E8E8E8}1.29 & \cellcolor[HTML]{E8E8E8}69.32 \\
 & gemma:7b & 600.35 & 38.60 & 3.58 & 51.04 & 616.57 & 39.67 & 3.57 & 72.95 \\
 & \cellcolor[HTML]{E8E8E8}granite-code:20b & \cellcolor[HTML]{E8E8E8}1015.81 & \cellcolor[HTML]{E8E8E8}69.11 & \cellcolor[HTML]{E8E8E8}8.26 & \cellcolor[HTML]{E8E8E8}27.89 & \cellcolor[HTML]{E8E8E8}1253.24 & \cellcolor[HTML]{E8E8E8}85.28 & \cellcolor[HTML]{E8E8E8}8.00 & \cellcolor[HTML]{E8E8E8}52.57 \\
 & granite-code:3b & 250.75 & 14.28 & 1.98 & 20.38 & 339.74 & 19.51 & 1.96 & 47.76 \\
 & \cellcolor[HTML]{E8E8E8}granite-code:8b & \cellcolor[HTML]{E8E8E8}464.86 & \cellcolor[HTML]{E8E8E8}30.63 & \cellcolor[HTML]{E8E8E8}3.57 & \cellcolor[HTML]{E8E8E8}27.61 & \cellcolor[HTML]{E8E8E8}528.69 & \cellcolor[HTML]{E8E8E8}35.02 & \cellcolor[HTML]{E8E8E8}3.55 & \cellcolor[HTML]{E8E8E8}48.16 \\
 & llama3:8b & 508.80 & 31.74 & 3.17 & 47.96 & 543.33 & 34.02 & 3.16 & 68.59 \\
 & \cellcolor[HTML]{E8E8E8}qwen2.5-coder:1.5b & \cellcolor[HTML]{E8E8E8}261.24 & \cellcolor[HTML]{E8E8E8}11.55 & \cellcolor[HTML]{E8E8E8}1.11 & \cellcolor[HTML]{E8E8E8}40.87 & \cellcolor[HTML]{E8E8E8}260.97 & \cellcolor[HTML]{E8E8E8}11.38 & \cellcolor[HTML]{E8E8E8}1.11 & \cellcolor[HTML]{E8E8E8}63.04 \\
 & qwen2.5-coder:7b & 512.73 & 32.58 & 3.17 & 47.94 & 461.37 & 28.87 & 3.15 & 68.52 \\
 & \cellcolor[HTML]{E8E8E8}qwen3:4b & \cellcolor[HTML]{E8E8E8}340.74 & \cellcolor[HTML]{E8E8E8}19.94 & \cellcolor[HTML]{E8E8E8}2.22 & \cellcolor[HTML]{E8E8E8}44.13 & \cellcolor[HTML]{E8E8E8}335.78 & \cellcolor[HTML]{E8E8E8}19.55 & \cellcolor[HTML]{E8E8E8}2.22 & \cellcolor[HTML]{E8E8E8}57.79 \\
\midrule
\multirow{16}{*}{4k} & \cellcolor[HTML]{E8E8E8}codegemma:7b & \cellcolor[HTML]{E8E8E8}733.47 & \cellcolor[HTML]{E8E8E8}47.70 & \cellcolor[HTML]{E8E8E8}3.80 & \cellcolor[HTML]{E8E8E8}52.95 & \cellcolor[HTML]{E8E8E8}713.41 & \cellcolor[HTML]{E8E8E8}46.23 & \cellcolor[HTML]{E8E8E8}3.79 & \cellcolor[HTML]{E8E8E8}\textbf{74.33} \\
 & codellama:13b & 1074.49 & 73.46 & 5.35 & 47.11 & 1031.47 & 70.56 & 5.35 & 62.29 \\
 & \cellcolor[HTML]{E8E8E8}codellama:7b & \cellcolor[HTML]{E8E8E8}624.59 & \cellcolor[HTML]{E8E8E8}42.26 & \cellcolor[HTML]{E8E8E8}3.01 & \cellcolor[HTML]{E8E8E8}47.88 & \cellcolor[HTML]{E8E8E8}597.12 & \cellcolor[HTML]{E8E8E8}40.31 & \cellcolor[HTML]{E8E8E8}3.00 & \cellcolor[HTML]{E8E8E8}65.44 \\
 & deepseek-coder-v2:16b & 639.20 & 30.01 & 1.75 & 44.88 & 528.30 & 24.55 & 1.66 & 62.58 \\
 & \cellcolor[HTML]{E8E8E8}deepseek-coder:1.3b & \cellcolor[HTML]{E8E8E8}209.20 & \cellcolor[HTML]{E8E8E8}11.98 & \cellcolor[HTML]{E8E8E8}0.94 & \cellcolor[HTML]{E8E8E8}38.95 & \cellcolor[HTML]{E8E8E8}198.94 & \cellcolor[HTML]{E8E8E8}11.30 & \cellcolor[HTML]{E8E8E8}0.93 & \cellcolor[HTML]{E8E8E8}53.49 \\
 & deepseek-coder:6.7b & 596.32 & 39.85 & 3.03 & 45.37 & 566.51 & 37.83 & 3.01 & 65.70 \\
 & \cellcolor[HTML]{E8E8E8}gemma:2b & \cellcolor[HTML]{E8E8E8}301.12 & \cellcolor[HTML]{E8E8E8}17.38 & \cellcolor[HTML]{E8E8E8}1.42 & \cellcolor[HTML]{E8E8E8}45.39 & \cellcolor[HTML]{E8E8E8}289.61 & \cellcolor[HTML]{E8E8E8}16.52 & \cellcolor[HTML]{E8E8E8}1.39 & \cellcolor[HTML]{E8E8E8}67.16 \\
 & gemma:7b & 717.37 & 46.52 & 3.76 & \textbf{53.27} & 703.91 & 45.55 & 3.73 & 72.83 \\
 & \cellcolor[HTML]{E8E8E8}granite-code:20b & \cellcolor[HTML]{E8E8E8}1386.40 & \cellcolor[HTML]{E8E8E8}94.53 & \cellcolor[HTML]{E8E8E8}8.77 & \cellcolor[HTML]{E8E8E8}28.22 & \cellcolor[HTML]{E8E8E8}1469.76 & \cellcolor[HTML]{E8E8E8}100.45 & \cellcolor[HTML]{E8E8E8}8.70 & \cellcolor[HTML]{E8E8E8}52.81 \\
 & granite-code:8b & 602.68 & 40.15 & 3.69 & 27.76 & 626.71 & 41.72 & 3.69 & 50.06 \\
 & \cellcolor[HTML]{E8E8E8}llama3:8b & \cellcolor[HTML]{E8E8E8}614.40 & \cellcolor[HTML]{E8E8E8}39.07 & \cellcolor[HTML]{E8E8E8}3.30 & \cellcolor[HTML]{E8E8E8}47.90 & \cellcolor[HTML]{E8E8E8}614.45 & \cellcolor[HTML]{E8E8E8}38.87 & \cellcolor[HTML]{E8E8E8}3.29 & \cellcolor[HTML]{E8E8E8}66.84 \\
 & qwen2.5-coder:1.5b & 308.24 & 13.43 & 1.14 & 40.51 & 301.51 & 12.80 & 1.11 & 60.81 \\
 & \cellcolor[HTML]{E8E8E8}qwen2.5-coder:7b & \cellcolor[HTML]{E8E8E8}602.72 & \cellcolor[HTML]{E8E8E8}38.63 & \cellcolor[HTML]{E8E8E8}3.25 & \cellcolor[HTML]{E8E8E8}46.05 & \cellcolor[HTML]{E8E8E8}539.13 & \cellcolor[HTML]{E8E8E8}34.09 & \cellcolor[HTML]{E8E8E8}3.23 & \cellcolor[HTML]{E8E8E8}66.10 \\
 & qwen3:4b & 409.62 & 24.32 & 2.29 & 42.26 & 383.46 & 22.56 & 2.26 & 59.27 \\
\midrule
\multirow{15}{*}{8k} & \cellcolor[HTML]{E8E8E8}codegemma:7b & \cellcolor[HTML]{E8E8E8}901.72 & \cellcolor[HTML]{E8E8E8}59.13 & \cellcolor[HTML]{E8E8E8}4.08 & \cellcolor[HTML]{E8E8E8}\textbf{56.14} & \cellcolor[HTML]{E8E8E8}870.22 & \cellcolor[HTML]{E8E8E8}56.78 & \cellcolor[HTML]{E8E8E8}4.08 & \cellcolor[HTML]{E8E8E8}\textbf{70.96} \\
 & codellama:13b & 1404.75 & 96.13 & 5.81 & 45.79 & 1387.39 & 94.89 & 5.88 & 61.77 \\
 & \cellcolor[HTML]{E8E8E8}codellama:7b & \cellcolor[HTML]{E8E8E8}841.48 & \cellcolor[HTML]{E8E8E8}56.94 & \cellcolor[HTML]{E8E8E8}3.32 & \cellcolor[HTML]{E8E8E8}51.75 & \cellcolor[HTML]{E8E8E8}826.93 & \cellcolor[HTML]{E8E8E8}55.89 & \cellcolor[HTML]{E8E8E8}3.38 & \cellcolor[HTML]{E8E8E8}66.08 \\
 & deepseek-coder-v2:16b & 848.72 & 39.24 & 1.76 & 45.67 & 790.12 & 36.68 & 1.83 & 59.28 \\
 & \cellcolor[HTML]{E8E8E8}deepseek-coder:1.3b & \cellcolor[HTML]{E8E8E8}297.69 & \cellcolor[HTML]{E8E8E8}17.44 & \cellcolor[HTML]{E8E8E8}1.12 & \cellcolor[HTML]{E8E8E8}37.32 & \cellcolor[HTML]{E8E8E8}296.44 & \cellcolor[HTML]{E8E8E8}17.37 & \cellcolor[HTML]{E8E8E8}1.12 & \cellcolor[HTML]{E8E8E8}48.12 \\
 & deepseek-coder:6.7b & 826.18 & 55.51 & 3.37 & 47.08 & 822.59 & 55.14 & 3.41 & 61.99 \\
 & \cellcolor[HTML]{E8E8E8}gemma:2b & \cellcolor[HTML]{E8E8E8}358.90 & \cellcolor[HTML]{E8E8E8}20.68 & \cellcolor[HTML]{E8E8E8}1.56 & \cellcolor[HTML]{E8E8E8}30.56 & \cellcolor[HTML]{E8E8E8}353.80 & \cellcolor[HTML]{E8E8E8}20.16 & \cellcolor[HTML]{E8E8E8}1.56 & \cellcolor[HTML]{E8E8E8}32.11 \\
 & gemma:7b & 892.09 & 58.43 & 4.05 & 53.30 & 868.27 & 56.53 & 4.07 & 69.27 \\
 & \cellcolor[HTML]{E8E8E8}granite-code:20b & \cellcolor[HTML]{E8E8E8}1958.13 & \cellcolor[HTML]{E8E8E8}133.17 & \cellcolor[HTML]{E8E8E8}10.00 & \cellcolor[HTML]{E8E8E8}31.60 & \cellcolor[HTML]{E8E8E8}2021.28 & \cellcolor[HTML]{E8E8E8}137.26 & \cellcolor[HTML]{E8E8E8}10.00 & \cellcolor[HTML]{E8E8E8}49.71 \\
 & llama3:8b & 765.83 & 49.11 & 3.53 & 47.34 & 750.91 & 48.03 & 3.50 & 63.99 \\
 & \cellcolor[HTML]{E8E8E8}qwen2.5-coder:1.5b & \cellcolor[HTML]{E8E8E8}359.49 & \cellcolor[HTML]{E8E8E8}16.56 & \cellcolor[HTML]{E8E8E8}1.26 & \cellcolor[HTML]{E8E8E8}38.44 & \cellcolor[HTML]{E8E8E8}361.44 & \cellcolor[HTML]{E8E8E8}16.15 & \cellcolor[HTML]{E8E8E8}1.25 & \cellcolor[HTML]{E8E8E8}58.84 \\
 & qwen2.5-coder:7b & 719.25 & 46.34 & 3.40 & 47.24 & 677.95 & 43.23 & 3.42 & 60.56 \\
 & \cellcolor[HTML]{E8E8E8}qwen3:4b & \cellcolor[HTML]{E8E8E8}496.04 & \cellcolor[HTML]{E8E8E8}29.59 & \cellcolor[HTML]{E8E8E8}2.38 & \cellcolor[HTML]{E8E8E8}44.53 & \cellcolor[HTML]{E8E8E8}474.78 & \cellcolor[HTML]{E8E8E8}28.03 & \cellcolor[HTML]{E8E8E8}2.39 & \cellcolor[HTML]{E8E8E8}58.72 \\
 \midrule
\multirow{10}{*}{12k} & \cellcolor[HTML]{E8E8E8}codellama:13b & \cellcolor[HTML]{E8E8E8}1876.01 & \cellcolor[HTML]{E8E8E8}128.58 & \cellcolor[HTML]{E8E8E8}6.54 & \cellcolor[HTML]{E8E8E8}43.60 & \cellcolor[HTML]{E8E8E8}1910.71 & \cellcolor[HTML]{E8E8E8}131.02 & \cellcolor[HTML]{E8E8E8}6.62 & \cellcolor[HTML]{E8E8E8}58.72 \\
 & codellama:7b & 1146.49 & 77.94 & 3.86 & \textbf{46.46} & 1170.43 & 79.50 & 3.92 & \textbf{62.24} \\
 & \cellcolor[HTML]{E8E8E8}deepseek-coder-v2:16b & \cellcolor[HTML]{E8E8E8}1113.96 & \cellcolor[HTML]{E8E8E8}52.34 & \cellcolor[HTML]{E8E8E8}2.10 & \cellcolor[HTML]{E8E8E8}42.22 & \cellcolor[HTML]{E8E8E8}1091.98 & \cellcolor[HTML]{E8E8E8}51.59 & \cellcolor[HTML]{E8E8E8}2.15 & \cellcolor[HTML]{E8E8E8}54.78 \\
 & deepseek-coder:1.3b & 433.49 & 25.99 & 1.40 & 34.51 & 448.06 & 26.80 & 1.42 & 42.56 \\
 & \cellcolor[HTML]{E8E8E8}deepseek-coder:6.7b & \cellcolor[HTML]{E8E8E8}1184.34 & \cellcolor[HTML]{E8E8E8}79.90 & \cellcolor[HTML]{E8E8E8}3.92 & \cellcolor[HTML]{E8E8E8}44.26 & \cellcolor[HTML]{E8E8E8}1215.03 & \cellcolor[HTML]{E8E8E8}82.01 & \cellcolor[HTML]{E8E8E8}3.98 & \cellcolor[HTML]{E8E8E8}56.88 \\
 & qwen2.5-coder:1.5b & 433.16 & 20.54 & 1.42 & 38.30 & 439.49 & 20.70 & 1.44 & 54.30 \\
 & \cellcolor[HTML]{E8E8E8}qwen2.5-coder:7b & \cellcolor[HTML]{E8E8E8}857.51 & \cellcolor[HTML]{E8E8E8}55.53 & \cellcolor[HTML]{E8E8E8}3.62 & \cellcolor[HTML]{E8E8E8}45.54 & \cellcolor[HTML]{E8E8E8}855.15 & \cellcolor[HTML]{E8E8E8}55.21 & \cellcolor[HTML]{E8E8E8}3.66 & \cellcolor[HTML]{E8E8E8}56.04 \\
 & qwen3:4b & 562.17 & 34.01 & 2.50 & 44.93 & 579.82 & 34.46 & 2.54 & 54.40 \\
\midrule
\multirow{10}{*}{16k} & \cellcolor[HTML]{E8E8E8}codellama:13b & \cellcolor[HTML]{E8E8E8}2357.74 & \cellcolor[HTML]{E8E8E8}161.85 & \cellcolor[HTML]{E8E8E8}7.25 & \cellcolor[HTML]{E8E8E8}41.79 & \cellcolor[HTML]{E8E8E8}2641.62 & \cellcolor[HTML]{E8E8E8}181.57 & \cellcolor[HTML]{E8E8E8}7.35 & \cellcolor[HTML]{E8E8E8}56.72 \\
 & codellama:7b & 1511.87 & 102.78 & 4.52 & 44.97 & 1651.18 & 112.54 & 4.46 & \textbf{59.93} \\
 & \cellcolor[HTML]{E8E8E8}deepseek-coder-v2:16b & \cellcolor[HTML]{E8E8E8}1326.67 & \cellcolor[HTML]{E8E8E8}63.53 & \cellcolor[HTML]{E8E8E8}2.42 & \cellcolor[HTML]{E8E8E8}42.76 & \cellcolor[HTML]{E8E8E8}1459.92 & \cellcolor[HTML]{E8E8E8}70.92 & \cellcolor[HTML]{E8E8E8}2.49 & \cellcolor[HTML]{E8E8E8}50.30 \\
 & deepseek-coder:1.3b & 600.05 & 36.04 & 1.69 & 32.23 & 667.24 & 40.36 & 1.74 & 41.98 \\
 & \cellcolor[HTML]{E8E8E8}deepseek-coder:6.7b & \cellcolor[HTML]{E8E8E8}1582.16 & \cellcolor[HTML]{E8E8E8}106.92 & \cellcolor[HTML]{E8E8E8}4.46 & \cellcolor[HTML]{E8E8E8}40.33 & \cellcolor[HTML]{E8E8E8}1738.78 & \cellcolor[HTML]{E8E8E8}117.83 & \cellcolor[HTML]{E8E8E8}4.55 & \cellcolor[HTML]{E8E8E8}55.11 \\
 & qwen2.5-coder:1.5b & 488.73 & 23.27 & 1.55 & 37.44 & 529.41 & 25.87 & 1.62 & 54.84 \\
 & \cellcolor[HTML]{E8E8E8}qwen2.5-coder:7b & \cellcolor[HTML]{E8E8E8}971.66 & \cellcolor[HTML]{E8E8E8}62.70 & \cellcolor[HTML]{E8E8E8}3.85 & \cellcolor[HTML]{E8E8E8}43.74 & \cellcolor[HTML]{E8E8E8}1038.28 & \cellcolor[HTML]{E8E8E8}67.33 & \cellcolor[HTML]{E8E8E8}3.94 & \cellcolor[HTML]{E8E8E8}54.55 \\
 & qwen3:4b & 637.32 & 37.94 & 2.62 & \textbf{45.16} & 686.45 & 40.88 & 2.70 & 55.23 \\
 \midrule
\multirow{4}{*}{24k} & \cellcolor[HTML]{E8E8E8}deepseek-coder-v2:16b & \cellcolor[HTML]{E8E8E8}1921.41 & \cellcolor[HTML]{E8E8E8}95.33 & \cellcolor[HTML]{E8E8E8}2.94 & \cellcolor[HTML]{E8E8E8}40.66 & \cellcolor[HTML]{E8E8E8}1993.03 & \cellcolor[HTML]{E8E8E8}98.77 & \cellcolor[HTML]{E8E8E8}2.90 & \cellcolor[HTML]{E8E8E8}53.94 \\
 & qwen2.5-coder:1.5b & 637.91 & 32.11 & 1.84 & 34.66 & 673.71 & 34.57 & 1.88 & 53.11 \\
 & \cellcolor[HTML]{E8E8E8}qwen2.5-coder:7b & \cellcolor[HTML]{E8E8E8}1275.13 & \cellcolor[HTML]{E8E8E8}83.22 & \cellcolor[HTML]{E8E8E8}4.26 & \cellcolor[HTML]{E8E8E8}\textbf{44.52} & \cellcolor[HTML]{E8E8E8}1368.26 & \cellcolor[HTML]{E8E8E8}89.59 & \cellcolor[HTML]{E8E8E8}4.27 & \cellcolor[HTML]{E8E8E8}\textbf{54.83} \\
 & qwen3:4b & 803.91 & 48.82 & 2.88 & 38.26 & 848.24 & 51.55 & 2.92 & 52.63 \\
 \midrule
\multirow{4}{*}{32k} & \cellcolor[HTML]{E8E8E8}deepseek-coder-v2:16b & \cellcolor[HTML]{E8E8E8}1816.94 & \cellcolor[HTML]{E8E8E8}91.11 & \cellcolor[HTML]{E8E8E8}3.41 & \cellcolor[HTML]{E8E8E8}40.58 & \cellcolor[HTML]{E8E8E8}2822.69 & \cellcolor[HTML]{E8E8E8}143.56 & \cellcolor[HTML]{E8E8E8}3.60 & \cellcolor[HTML]{E8E8E8}45.90 \\
 & qwen2.5-coder:1.5b & 679.48 & 32.80 & 2.00 & 35.49 & 911.62 & 48.67 & 2.26 & 44.92 \\
 & \cellcolor[HTML]{E8E8E8}qwen2.5-coder:7b & \cellcolor[HTML]{E8E8E8}1322.70 & \cellcolor[HTML]{E8E8E8}85.11 & \cellcolor[HTML]{E8E8E8}4.59 & \cellcolor[HTML]{E8E8E8}42.15 & \cellcolor[HTML]{E8E8E8}1874.55 & \cellcolor[HTML]{E8E8E8}123.37 & \cellcolor[HTML]{E8E8E8}4.81 & \cellcolor[HTML]{E8E8E8}\textbf{48.11} \\
 & qwen3:4b & 813.82 & 48.39 & 3.00 & \textbf{46.41} & 1083.22 & 66.65 & 3.21 & 46.93 \\
\bottomrule
\end{tabular}
\end{threeparttable}
\end{table*}

\subsubsection{RQ1.1-How does the energy-accuracy trade-off vary across programming languages?}

To further investigate whether the observed trade-offs differ across programming languages, we performed statistical analyses on the achieved accuracy scores. 
In the McEval dataset (Table~\ref{tab:mc_res}), applying Friedman's test revealed a significant effect of programming language on Pass@1 ($\chi^2 = 40.36$, $p < 0.001$). Post-hoc Wilcoxon signed-rank tests with Bonferroni correction ($\alpha = 0.0167$) showed no significant difference between Python and Java ($p = 0.90$). However, both Python and Java significantly outperformed Rust ($p < 0.001$), with median paired differences of 5.75 and 5.01, respectively. We did not perform similar tests for energy and time because the workloads differ across programming languages. 

In the RepoBench dataset (Table~\ref{tab:repo_res}), we report Edit Similarity (ES) as a proxy for accuracy. A Wilcoxon signed-rank test reveals a statistically significant difference between Python and Java ($W = 103.0$, $p < 0.001$). Java achieved a higher median ES (56.43) compared to Python (43.02), with a median paired difference of $-14.90$.
One possible explanation for these findings lies in the characteristics of the programming languages. Python, being dynamically typed, and Java, being statically typed, may influence how LLMs perform under syntax-based evaluation metrics such as ES. In particular, the stricter and more explicit structure of Java code could partially explain its higher ES scores. However, when evaluated using execution-based metrics such as Pass@1 (as in McEval), the performance of Python and Java becomes comparable.
In contrast, Rust is consistently outperformed by both Python and Java. This may be due to the relatively smaller amount of Rust code available in the training corpora of contemporary LLMs.

As shown in Figure~\ref{fig:mc_pareto}, in McEval, both quantized variants of the small \texttt{Qwen2.5-coder:1.5b} model are located close to the ideal point while still achieving competitive Pass@1 scores across all three languages. This indicates that smaller quantized models can provide a favorable balance between energy efficiency and accuracy.  
Among the larger models, the 4-bit quantized variant of \texttt{Deepseek-coder-v2:16b} stands out as one of the most efficient high-performing models. Despite containing 16B total parameters, this MoE model activates only approximately 4B parameters during inference, allowing it to achieve the highest Pass@1 scores in Java and Rust while remaining on the Pareto frontier. In Python, however, \texttt{Qwen2.5-coder:7b-q4} achieves the highest accuracy overall. The improvement over \texttt{Deepseek-coder-v2:16b-q4} is approximately 16\% while requiring nearly twice as much energy per generated token.

Notably, the Pareto frontiers across languages are largely composed of the same model families, suggesting that the energy-accuracy trade-off remains relatively stable across programming languages. One exception is \texttt{Qwen2.5-coder:7b}, which performs particularly well in Python but is less competitive in Java and Rust. Most Pareto-optimal configurations correspond to quantized models. Even when FP16 models appear on the frontier, their quantized counterparts often achieve nearly identical accuracy while consuming substantially less energy per generated token. Several larger FP16 models are dominated by smaller quantized alternatives, indicating that increasing parameter count and precision does not necessarily yield better energy-accuracy trade-offs. Overall, quantized models frequently provide the best balance between accuracy and energy efficiency, with small accuracy degradation relative to their energy savings.

\subsubsection{RQ1.2-How does the energy-accuracy trade-off vary across different context sizes?}
\review{In RepoBench, as shown in Table~\ref{tab:repo_res}, increasing the context size generally leads to a higher energy cost per generated token (J/token) in both Python and Java. For example, in Python, \texttt{Qwen2.5-coder:1.5b} increases from 1.11 J/token at a context size of 2000 to 2.00 J/token at 32000, with a similar trend observed in Java. This pattern is reflected in Figure~\ref{fig:repo_pareto}, where models gradually shift toward higher J/token values as the context size increases. This trend is further summarized across models in the online appendix~\cite{anonymous_replication_2026}.}

Although the workloads across different context sizes are not identical in terms of both content and size, they are designed to represent repository-level completion tasks of comparable complexity. We found out that increasing the context size does not consistently translate into substantial improvements in ES, suggesting diminishing returns from larger context windows.
For some models, such as \texttt{Granite-code:20b} and \texttt{Gemma:7b}, ES slightly improves as larger contexts are provided. In contrast, for models such as \texttt{Codellama:7b} and \texttt{Codellama:13b}, performance gradually decreases with increasing context size. Other models, including \texttt{Qwen2.5-coder:1.5b} and \texttt{Qwen2.5-coder:7b}, exhibit modest improvements up to a certain context length, after which ES begins to decline.

Notably, the composition of the Pareto frontier remains relatively stable across context sizes, with quantized variants frequently appearing among the Pareto-optimal configurations. This confirms that increasing model precision or context size does not necessarily result in proportionally better quality also in repository-level code completion. In several cases, quantized variants achieve nearly identical ES compared to their FP16 counterparts while consuming substantially less energy per generated token.

Model rankings were not stable in terms of per-token energy consumption across context sizes. In the models with size lower than 2B range, \texttt{Deepseek-coder:1.3b} exhibited lower J/token values than \texttt{Qwen2.5-coder:1.5b} up to a context size of 12K. However, at 16K context, \texttt{Qwen2.5-coder:1.5b} became more energy-efficient. A similar trend was observed in the 4B range between \path{Deepseek-coder-v2:16b-lite-base-fp16} and \texttt{Qwen3:4b-fp16}, both of which have approximately 4B active parameters. While the DeepSeek model remained more efficient at smaller contexts, \texttt{Qwen3:4b-fp16} achieved lower J/token values at 32K context. Likewise, among the 7B models, \texttt{Qwen2.5-coder:7b-base-fp16} became more energy-efficient than both \texttt{Codellama:7b-code-fp16} and \path{Deepseek-coder:6.7b-base-fp16} starting from 12K context size. These observations suggest that certain model families may exhibit better scalability under long-context workloads.

\begin{summary}
{
\textbf{Summary.} 
The trade-off between energy consumption and accuracy of LLMs is not always strict. In code completion, it is strongly influenced by the characteristics of the completion task, programming language, and context size. Increasing context size consistently raises energy consumption, while the corresponding accuracy improvements are often limited. Therefore, no single model consistently provides the best trade-off across all tasks, programming languages, and context sizes.

\review{\textbf{Takeaway.} Practitioners should select models based on the programming language and expected context length of their target use case rather than a model’s reported accuracy alone. Furthermore, model providers should report inference energy alongside accuracy.}}
\end{summary}

\subsection{RQ2: What is the relative influence of input and output token counts on total energy consumption, and how do these drivers interact with model scale?}
\begin{table}[ht]\centering
\caption{Spearman correlations between total energy consumption and the studied predictors, including input tokens, output tokens, active parameter count, and their interaction terms. Bonferroni correction was applied ($\alpha = 0.05/6 = 0.0083$) and all correlations remain significant.}
\label{tab:correlation_results}
\renewcommand{\arraystretch}{0.95}
\small
\begin{tabular}{lcc}
\toprule
\textbf{Predictor} & \textbf{RepoBench} & \textbf{McEval} \\
\midrule
Input Tokens & 0.57 & 0.24 \\
Output Tokens & 0.40 & 0.36 \\
Active Parameter Count & 0.60 & 0.75 \\
\midrule
Input Tokens $\times$ Active Parameter Count & 0.94 & 0.79 \\
Output Tokens $\times$ Active Parameter Count & 0.67 & 0.92 \\
Input Tokens $\times$ Output Tokens & 0.58 & 0.40 \\
\bottomrule
\end{tabular}
\end{table}
To quantify the drivers of energy consumption, we first analyzed the pairwise relationships between energy usage, token counts (input and output), and active parameter count using correlation analysis. The results are presented in Table~\ref{tab:correlation_results}. We selected input and output token counts because both prompt processing and autoregressive generation contribute directly to inference cost. Active parameter count is a compact proxy for model scale and computational complexity, as larger models consistently consumed more energy than smaller ones under comparable workloads. For McEval, correlation was high for active parameter count and weak for the other two variables. This reflects the characteristics of this dataset: input and output lengths are short, thus model scale dominates energy use. For RepoBench, output length is short, and input length is very long, so their impact is comparable to the model scale. In this case, energy use is moderately correlated with the three variables. 

In addition, we included interaction terms between token counts and active parameter count to capture the intuition that the computational cost of processing or generating tokens depends not only on the number of tokens, but also on the size of the model performing the computation. Moreover, the interaction between input and output tokens was included to account for the fact that autoregressive generation depends on the size of the provided context. This effect is also reflected in the RepoBench Pareto plots (Figure~\ref{fig:repo_pareto}), where models consistently require higher energy per generated token as context size increases, despite the output length being capped at 128 tokens. This suggests that generating the same number of output tokens may incur substantially different costs under short and long context workloads. As shown in Table~\ref{tab:correlation_results}, the first two interaction variables exhibit strong correlation with energy consumption across both benchmarks, while the interaction between input and output tokens is moderately correlated.  
These interaction terms are particularly relevant in our setting, since RepoBench emphasizes long-context processing whereas McEval contains generation-heavy workloads.
To further analyze the contribution of these factors, we fitted multiple linear regression models using standardized variables ($\mu=0, \sigma=1$). \review{As input token count, output token count, and active parameter count are measured on different scales, we standardized these predictors to enable direct comparison of their regression coefficients.} The regression assumptions were reasonably satisfied, and the corresponding diagnostic plots are provided in the online appendix~\cite{anonymous_replication_2026}.
The general regression model is defined in Equation~\ref{eq:regression_model}.
\begin{equation}
E = \beta_0 + \beta_1 I + \beta_2 O + \beta_3 P +
\beta_4 I \cdot O +
\beta_5 I \cdot P +
\beta_6 O \cdot P +
\beta_7 W +
\beta_8 L +
\varepsilon
\label{eq:regression_model}
\end{equation}
\noindent where $E$ denotes total energy consumption (Wh), $I$ and $O$ represent the standardized input and output token counts, $P$ is the standardized active parameter count, $W$ denotes bit width (quantization), $L$ represents the programming language, and $\varepsilon$ is the error term. The interaction terms capture that the energy cost of generation depends jointly on context size and model scale.
To facilitate interpretation of the regression coefficients, the standard deviations (SD) of the predictors are reported in Table~\ref{tab:std_values}.
The dependent variable is the total energy consumption in Wh. As predictors, we included standardized input tokens, output tokens, and active parameter count, as well as interaction terms between these variables. Bit width and programming language were treated as categorical variables. To account for multiple observations from the same model families, we utilized Cluster-Robust Standard Errors, grouping by model families. This ensures our $p$-values are not artificially inflated by intra-model correlation. \review{For RepoBench, the statistics and regressions are computed using the common models evaluated across all context sizes up to 16K to ensure consistent model coverage, as only a few models support 24K and 32K contexts. For McEval, all evaluated models are included.} The regression results are presented in Table~\ref{tab:regression_results}.
\begin{table}[h]\centering
\caption{Standard deviations of the standardized predictors used in the regression models. Token statistics represent the aggregated number of tokens across all prompts within each benchmark.}
\label{tab:std_values}
\renewcommand{\arraystretch}{0.95}
\small
\begin{tabular}{lccc}
\toprule
\textbf{Benchmark} & \textbf{Input Tokens} & \textbf{Output Tokens} & \textbf{Parameter Count} \\
\midrule
RepoBench & 1,316,930.93 & 2,309.96 & 4,563.42 \\
McEval & 21,161.11 & 39,171.61 & 4,781.41 \\
\bottomrule
\end{tabular}
\end{table}
\begin{table*}[!htp]\centering
\caption{Regression coefficients ($\beta$), standard errors (SE), statistical significance ($p$-value), and model fit metrics for RepoBench and McEval. Cluster-robust standard errors are used (clustered by family). Q4 is used as the reference category for bit width, and Java is the reference language in both RepoBench and McEval.}

\label{tab:regression_results}
\renewcommand{\arraystretch}{0.95}
\scriptsize
\begin{tabular}{lrrrrr}
\toprule
\multicolumn{6}{c}{\textbf{RepoBench}} \\
\midrule
\textbf{Predictor} & \textbf{coef ($\beta$)} & \textbf{SE} & \textbf{$p$-value} & \textbf{$R^2$} & \textbf{Adj. $R^2$} \\
\midrule
Intercept & 49.285 & 1.889 & $<$0.001 & \multirow{10}{*}{0.961} & \multirow{10}{*}{0.960} \\

Bit width (Q8) & 2.389 & 1.059 & 0.024 & & \\
Bit width (FP16) & 3.008 & 1.014 & 0.003 & & \\

Language (Python) & 1.433 & 0.535 & 0.007 & & \\

Input Tokens$_{std}$ & 22.976 & 1.431 & $<$0.001 & & \\
Output Tokens$_{std}$ & 4.902 & 1.916 & 0.011 & & \\
Active Parameter Count$_{std}$ & 26.604 & 1.497 & $<$0.001 & & \\

Input Tokens$_{std}$ $\times$ Output Tokens$_{std}$ & 3.978 & 1.583 & 0.012 & & \\
Input Tokens$_{std}$ $\times$ Active Parameter Count$_{std}$ & 9.788 & 1.328 & $<$0.001 & & \\
Output Tokens$_{std}$ $\times$ Active Parameter Count$_{std}$ & 1.448 & 0.987 & 0.142 & & \\

\midrule
\multicolumn{6}{c}{\textbf{McEval}} \\
\midrule
\textbf{Predictor} & \textbf{coef ($\beta$)} & \textbf{SE} & \textbf{$p$-value} & \textbf{$R^2$} & \textbf{Adj. $R^2$} \\
\midrule
Intercept & 40.748 & 3.718 & $<$0.001 & \multirow{11}{*}{0.940} & \multirow{11}{*}{0.937} \\

Bit width (Q8) & 5.841 & 1.232 & $<$0.001 & & \\
Bit width (FP16) & 21.947 & 3.027 & $<$0.001 & & \\

Language (Python) & -0.030 & 6.149 & 0.996 & & \\
Language (Rust) & 0.058 & 1.434 & 0.968 & & \\

Input Tokens$_{std}$ & 0.427 & 2.887 & 0.882 & & \\
Output Tokens$_{std}$ & 23.496 & 2.189 & $<$0.001 & & \\
Active Parameter Count$_{std}$ & 27.259 & 1.171 & $<$0.001 & & \\

Input Tokens$_{std}$ $\times$ Output Tokens$_{std}$ & -0.352 & 0.209 & 0.093 & & \\
Input Tokens$_{std}$ $\times$ Active Parameter Count$_{std}$ & -0.562 & 0.394 & 0.154 & & \\
Output Tokens$_{std}$ $\times$ Active Parameter Count$_{std}$ & 15.674 & 1.351 & $<$0.001 & & \\
\bottomrule
\end{tabular}
\end{table*}

\noindent
For RepoBench, the regression model achieved a high $R^2$ (0.961), indicating that the selected predictors explain most observed variation in energy consumption. Input token count and active parameter count exhibit the strongest positive associations with energy usage, with coefficients of 22.98 and 26.60, respectively. In contrast, output token count has a smaller but statistically significant effect. This is expected, as repository-level context can grow substantially while generation length remains capped at 128 tokens to predict the next line.
\\
Because the predictors were standardized, we can directly compare the regression coefficients to measure the real-world impact of scaling up data sizes. The coefficients represent the expected change in total energy consumption (in Wh) for every \review{one-SD} increase in that variable. For example, in RepoBench, a \review{one-SD} increase in input tokens corresponds to approximately 1.3 million (1,316,930) additional aggregated input tokens across the 300 prompt instances ($\sim 4333$ tokens per prompt), which is associated with an increase of approximately 23 Wh in total energy consumption.
On the other hand, a one-SD increase in output tokens (2310 tokens), imposes 4.9 Wh more energy in total energy usage. 

To find out how many output tokens would generate that same 22.98 Wh ($\sim 23$) increase, we can calculate their energy equivalence. While a 23 Wh increase in energy consumption is driven by an additional 1.3 million input tokens, that exact same 23 Wh energy increase would be triggered by generating just 10,857 aggregated output tokens ($\sim 36$ token per prompt). Under the RepoBench workload configuration, generating a single token consumes as much energy as processing approximately 121 input tokens ($1,316,930 / 10,857$).

The interaction between input tokens and active parameter count is also statistically significant, indicating that the energy impact of increasing context size becomes larger for bigger models. Similarly, the interaction between input and output tokens is significant, suggesting that generation becomes more expensive when conditioned on larger contexts. However, the interaction between output tokens and parameter count is not statistically significant, likely because the variation in generated output length is relatively limited in RepoBench compared to the variation in input size.

Bit width also has a statistically significant effect on energy consumption. Compared to the Q4 baseline, Q8 and FP16 variants require additional energy, confirming that quantization (lower bit width) can improve energy efficiency even after controlling for token counts and parameter count. Moreover, the programming language variable is statistically significant, suggesting that language-specific characteristics such as tokenization patterns and repository structure may influence energy consumption in long-context workloads.

For McEval, the regression model also achieved a high goodness of fit ($R^2 = 0.940$). However, the contribution of the predictors differs substantially from RepoBench. In this benchmark, input size is capped at approximately 2K tokens, while the generation length can reach up to 1300 tokens. As a result, output token count becomes one of the dominant predictors of energy consumption, with a coefficient of 23.50, whereas input token count is no longer statistically significant. This indicates that, in generation-heavy workloads, the decoding phase contributes more strongly to total energy usage than prompt processing.

The interaction between output tokens and parameter count is highly significant in McEval, indicating that generating longer outputs becomes increasingly expensive for larger models. In contrast, the interactions involving input tokens are not statistically significant, which aligns with the relatively limited variation in prompt length in this benchmark. Quantization effects remain statistically significant, with FP16 models showing substantially higher energy consumption than Q4 variants. However, unlike RepoBench, the programming language variables are not statistically significant after controlling for token counts and parameter count, suggesting that language-specific effects become less influential when the workload is dominated by generation rather than long-context processing. The regression behavior closely matches the intended workload characteristics of the two benchmarks.

These findings align with the earlier correlation analysis, where interaction terms exhibited the strongest associations with energy consumption in their respective code completion workload settings.
Overall, the two benchmarks provide complementary perspectives on the energy behavior of LLM inference. RepoBench primarily captures the effect of long-context processing, where input size and its interaction with parameter count dominate energy consumption. In contrast, McEval emphasizes generation-heavy workloads, where output length and its interaction with parameter count become the primary drivers. Together, these results indicate that the dominant contributors to energy consumption depend strongly on the structure of the workload. Context-heavy software engineering tasks are mainly influenced by prompt processing costs, whereas generation-heavy tasks are primarily driven by autoregressive decoding behavior.

\begin{summary}
{
\textbf{Summary.} 
Our regression analysis shows that the dominant drivers of energy consumption strongly depend on the characteristics of the completion task. In RepoBench, energy usage is primarily influenced by input context size and its interaction with active parameter count, highlighting the cost of long-context processing. In contrast, McEval is mainly driven by output generation, its interaction with active parameter count, and quantization level, indicating that autoregressive decoding dominates in generation-heavy workloads.
\\
The results also reveal that output generation is substantially more energy-intensive on a per-token basis than prompt processing. Furthermore, the significance of the interaction terms suggests that the energy cost of processing and generating tokens is better explained by considering token counts and the scale of the model together.

\review{\textbf{Takeaway.} Practitioners should select models carefully, as larger models do not necessarily yield better accuracy but can considerably increase energy consumption. Furthermore, limiting unnecessary input and output tokens can reduce inference energy.}
}
\end{summary}

\section{Related Work}
Recent research on LLMs for software engineering (SE) focuses on improving code completion accuracy, while the inference energy footprints remain underexplored. A systematic review~\cite{verdecchia2023systematic} shows that 76\% of Green AI literature was published recently and nearly half of these primary studies focus only on the training phase. Most remaining inference benchmarks are algorithm-agnostic or use image datasets. This leaves open questions about how inference energy scales across programming languages, context windows, and model precisions.

Several studies have investigated LLM-based code completion from an accuracy perspective, including efficient training strategies for fill-in-the-middle (FIM) models, syntax-aware infilling benchmarks, practical IDE-based evaluation settings, alignment techniques for improving code generation quality, and empirical analyses of LLM-generated software quality~\cite{bavarian2022efficient,gong2024eval,izadi2024language,ren2025alignment,molison2025llm}. These studies provide valuable insights into code completion performance and software quality, but they do not primarily analyze the energy cost of inference.

Repository-level code completion has also received increasing attention, since real-world code completion often depends on information across multiple files. Recent studies emphasize long-context reasoning and cross-file dependency modeling through retrieval augmentation and contextual integration strategies~\cite{liu2024graphcoder,guan2024contextmodule, hu2026line}. Zhang et al.~\cite{zhang2023repocoder} proposed RepoCoder, which combines retrieval and generation for repository-level code completion, while Ding et al.~\cite{ding2024cocomic} studied code completion by jointly modeling in-file and cross-file context. Recent work extends repository-level completion to specific domains, such as RTL code completion~\cite{wu2025rtlrepocoder}, and API-centric repository-level code completion~\cite{li2025apirepo}. These studies emphasize repository context and retrieval, but they mainly evaluate completion quality rather than the energy implications of processing larger contexts.

Beyond accuracy, related work studies sustainability in code generation. Some focus on the energy efficiency of the generated code itself, examining whether LLM-generated programs are more or less energy-efficient than human-written solutions~\cite{cursaru2024controlled, vartziotis2024learn, islam2025evaluating, huang2024effibench, apsan2025generating}. This work is complementary to ours: it studies the runtime energy behavior of the produced software, whereas our study focuses on the energy usage of the LLM during the code completion.

Energy consumption of LLMs has also been studied from Green AI and inference-efficiency perspectives. Fernandez et al.~\cite{fernandez2025energy} analyze the energy implications of LLM inference optimizations across diverse workloads, software stacks, hardware accelerators, and serving configurations. GreenMyLLM~\cite{coignion2024green} studies the energy consumption of code assistants by simulating developer interactions and analyzing factors such as model size, quantization, streaming, and rejected suggestions. Jareno et al.~\cite{JARENO2026108483} further investigate energy-efficient LLM inference from an SE perspective, proposing compiler-level optimization of the \textit{llama.cpp} inference engine and showing that hardware-aware code transformations can reduce energy consumption without modifying the underlying model. More recently, SweetSpot~\cite{pizzini2026sweetspot} proposed an analytical model for predicting LLM inference energy efficiency as a function of input and output sequence lengths, highlighting non-linear efficiency regimes and energy “sweet spots” under controlled synthetic workloads. These works demonstrate the growing importance of inference energy, but they do not focus specifically on comparing realistic code completion workloads across context sizes, programming languages, and completion styles.

The studies closest to ours jointly consider LLM inference energy and SE tasks. Solovyeva and Castor~\cite{solovyeva2026towards} perform a phase-level analysis of LLM inference energy in software development benchmarks, distinguishing between input processing and output generation and showing that excessive generation can substantially increase energy consumption. Mehditabar et al.~\cite{mehditabar2025smart} propose BRACE, a benchmark-oriented framework for evaluating coding LLMs in terms of both functional accuracy and energy efficiency. These studies are highly relevant, but they address different levels of analysis. The former focuses on phase-level inference behavior, while the latter provides a unified energy-accuracy benchmarking perspective.

Our work complements these studies by examining workload-level energy behavior in realistic code completion scenarios. We study both repository-level left-to-right completion with varying context sizes and multilingual fill-in-the-middle completion across Python, Java, and Rust. Rather than only ranking models by accuracy or energy, we analyze how input length, output length, active parameter count, quantization, programming language, and context size shape energy-accuracy trade-offs. We show that long-context repository completion and generation-heavy FIM completion stress LLM inference differently, and that energy-efficient model selection depends on the characteristics of the completion task.

\section{Threats to Validity}
A potential threat to \textbf{internal validity} relates to the occasional incomplete responses returned by the Ollama API. In a small number of cases (typically one or two prompts per model), the model failed to return a complete generation and skipped the query. This issue became more noticeable in RepoBench under large context sizes (e.g., beyond 16K tokens), where missing a single long-running prompt could noticeably reduce the measured elapsed time and energy consumption, potentially making some FP16 models appear more energy-efficient than lower-bit quantized variants. To mitigate this issue, we manually inspected the results and excluded incomplete executions when computing the energy cost per generated token.
Another threat to internal validity arises from the limited number of repetitions per experimental configuration. Each setup was executed three times to capture transient runtime variability while keeping the overall computational cost manageable. Increasing the number of repetitions could further improve statistical confidence, but would significantly increase the required execution time and resource usage. Nevertheless, since each run processes hundreds of prompts over several minutes and GPU power is sampled at 10Hz, the measurements exhibit relatively stable utilization patterns, reducing the impact of transient fluctuations.
Another threat concerns the approximation used for energy estimation. Total energy consumption was computed using the arithmetic mean of the sampled GPU power values multiplied by the elapsed execution time, rather than performing numerical integration over the full power trace. To assess the impact of this approximation, we conducted an additional validation experiment on the Java subset of RepoBench. The observed discrepancy between the two approaches remained small, with an average difference of 0.46\% and a maximum deviation of 2.3\%.
\review{Finally, energy was measured using a software-based GPU power profiler rather than external hardware instruments, which may affect measurement precision. However, the same procedure was applied consistently across all configurations.}

A potential threat to \textbf{external validity} of our study concerns the selection of programming languages. Although Python, Java, and Rust cover dynamically typed, statically typed, and relatively lower-resource mainstream languages, the results may not generalize to all programming ecosystems or domain-specific languages.
Another potential threat to external validity is that the measurements on the A100 platform are limited to GPU power obtained through NVML, since privileged access required for CPU-side energy measurements was unavailable on the cluster. Consequently, the reported energy values do not include the contribution of CPU, DRAM, or PCIe communication overheads. Although this may lead to an underestimation of total system-level energy consumption, all experiments were conducted under identical hardware configurations and comparable data transfer conditions, which helps preserve the validity of fair comparisons across models and workloads.

\review{A potential threat to \textbf{conclusion validity} concerns regression assumptions. Diagnostic Q-Q plots exhibit deviations from normality in the residual tails. Alternative specifications using a log-transformed outcome and Gamma regression did not consistently improve the diagnostics. Therefore, we acknowledge these deviations as a limitation of the analysis.}

\section{Conclusion}
This study investigated the trade-off between accuracy and energy consumption in LLM-based code completion across different programming languages, context sizes, and completion workloads. Using RepoBench and McEval, we evaluated \review{a total of 25} open-weight LLMs under long-context repository-level completion and generation-heavy fill-in-the-middle completion scenarios.
Our results show that the dominant drivers of energy consumption depend strongly on the characteristics of the completion task. In long-context workloads such as RepoBench, energy consumption is primarily influenced by input context size and its interaction with model scale. In contrast, generation-heavy workloads such as McEval are mainly driven by output generation and its interaction with active parameter count. These findings indicate that different software engineering tasks stress different components of LLM inference.
\\
We further observed that output generation is substantially more energy-intensive per token than prompt processing. In addition, increasing context size consistently raised the energy cost per generated token, while the corresponding improvements in completion quality were often limited and highly model-dependent. Across both benchmarks, smaller and heavily quantized models frequently achieved Pareto-optimal trade-offs, often providing accuracy comparable to larger full precision models while consuming substantially less energy. Moreover, model rankings in terms of energy efficiency were not stable across context sizes, suggesting that certain model families scale more efficiently under long-context workloads.
\\
Overall, our findings demonstrate that larger models and longer contexts do not necessarily provide proportionally better completion quality, emphasizing the importance of energy-aware model selection and deployment strategies for AI-assisted software engineering tools.

Future work could complement our regression analysis with controlled ablation studies that vary one factor at a time while keeping others fixed, enabling a more precise understanding of how context length, output length, quantization, and model scale influence energy consumption.
In addition, extending this analysis to emerging agentic workflows involving iterative multi-turn interactions, self-correction, and tool usage would help better understand the sustainability challenges of large-scale AI-assisted software engineering systems.
\section{Data Availability}
The replication package for this study, along with the appendices, is publicly available~\cite{anonymous_replication_2026}.

\bibliographystyle{plainurl}
\bibliography{references.bib}

\end{document}